%% file: main.tex
\documentclass[conference]{IEEEtran}
\ifCLASSINFOpdf
\else
\fi
\usepackage{epsfig,endnotes,amssymb,makecell,hyperref}
\usepackage{subcaption,threeparttable,booktabs,multirow,xurl}
\usepackage{float}
\usepackage{stfloats} % or use \usepackage{dblfloatfix}
\usepackage{ulem}
\usepackage[most]{tcolorbox}
\usepackage[linesnumbered,ruled,vlined]{algorithm2e}
\usepackage{pifont}

\usepackage{listings}
\usepackage{xcolor}
\usepackage{soul}         % for \hl
\usepackage{caption}
\usepackage{setspace}
\usepackage{makecell}

\definecolor{lightgray}{gray}{0.95}
\definecolor{myred}{rgb}{0.8,0.1,0.1}
\definecolor{myblue}{rgb}{0.1,0.1,0.7}

\usepackage[utf8]{inputenc}
\usepackage[T1]{fontenc}

\newif\ifcommentcond
\commentcondtrue % enable comments
\newcounter{wqy} % wqy

\newcounter{pgn} % pgn

\begin{document}
\title{Breaking Isolation: A New Perspective on Hypervisor Exploitation via Cross-Domain Attacks }

\author{
    \IEEEauthorblockN{
        Gaoning Pan\IEEEauthorrefmark{1}\IEEEauthorrefmark{2},
        Yiming Tao\IEEEauthorrefmark{3},
        Qinying Wang\IEEEauthorrefmark{4}\IEEEauthorrefmark{3},
        Chunming Wu\IEEEauthorrefmark{3},
        Mingde Hu\IEEEauthorrefmark{1}\IEEEauthorrefmark{2},
        Yizhi Ren\IEEEauthorrefmark{1}\IEEEauthorrefmark{2}\textsuperscript{\(\star\)}, 
        Shouling Ji\IEEEauthorrefmark{3}
    }
    
    \IEEEauthorblockA{
        % 第一行：放置所有的“大学”单位 (1, 3, 4)，它们较短，可以并排
        \IEEEauthorrefmark{1}Hangzhou Dianzi University, \quad
        \IEEEauthorrefmark{3}Zhejiang University, \quad
        \IEEEauthorrefmark{4}EPFL
    }
    
    \IEEEauthorblockA{
        % 第二行：放置最长的“实验室”单位 (2)，让它独占一行居中
        \IEEEauthorrefmark{2}Zhejiang Provincial Key Laboratory of Sensitive Data Security and Confidentiality Governance
    }
    
    \IEEEauthorblockA{
        % 第三行：邮箱合并
        \textit{Email: \{pgn, 20227001, renyz\}@hdu.edu.cn, \{taoym, wuchunming, sji\}@zju.edu.cn, qinying.wang@epfl.ch}
    }

    \thanks{
        \(\star\) Corresponding Author: Yizhi Ren (renyz@hdu.edu.cn)
    }
}


\IEEEoverridecommandlockouts
\makeatletter\def\@IEEEpubidpullup{6.5\baselineskip}\makeatother
\IEEEpubid{\parbox{\columnwidth}{
		Network and Distributed System Security (NDSS) Symposium 2026\\
		23 - 27 February 2026 , San Diego, CA, USA\\
		ISBN 979-8-9919276-8-0\\
		https://dx.doi.org/10.14722/ndss.2026.240376\\
		www.ndss-symposium.org
}
\hspace{\columnsep}\makebox[\columnwidth]{}}

\maketitle

\begin{abstract}

% 1. 数据结构稀疏，高度定制，搜索困难。现有漏洞利用工具搜索效率低
% 2. 数据结构定位困难，存在ASLR，利用时候依赖条件高（需要0-day 信息泄漏）

Hypervisors are under threat by critical memory safety vulnerabilities, with pointer corruption being one of the most prevalent and severe forms. Existing exploitation frameworks depend on identifying highly-constrained structures in the host machine and accurately determining their runtime addresses, which is ineffective in hypervisor environments where such structures are rare and further obfuscated by Address Space Layout Randomization (ASLR). We instead observe that modern virtualization environments exhibit \textit{weak memory isolation} — guest memory is fully attacker-controlled yet accessible from the host, providing a reliable primitive for exploitation. Based on this observation, we present the first systematic characterization and taxonomy of Cross-Domain Attacks (CDA), a class of exploitation techniques that enable capability escalation through guest memory reuse. To automate this process, we develop a system that identifies cross-domain gadgets, matches them with corrupted pointers, synthesizes triggering inputs, and assembles complete exploit chains. Our evaluation on 15 real-world vulnerabilities across QEMU and VirtualBox shows that CDA is widely applicable and effective.

\end{abstract}
% \input{0x00Abstract}


% \fi
\IEEEpeerreviewmaketitle

\input{0x01Introduction}
\input{0x02Background}

\input{0x03Motivation}

\input{0x04Methodology}

\input{0x05Evaluation}
\input{0x06Discussion}
\input{0x07RelatedWork}

\input{0x08Conclusion}

\section*{Ethics Consideration}
In conducting our research, we focus exclusively on known hypervisor vulnerabilities, all of which have already been patched by vendors. Furthermore, all experiments are conducted in controlled local environments, fully isolated from any public cloud infrastructure. As such, our work does not pose any risk to production systems or real-world deployments, and does not raise any ethical concerns. In addition, we have proactively communicated our findings to the QEMU development team to discuss the broader implications of CDA and potential mitigation strategies. While the developers are still assessing the long-term impact, we have proposed several mitigation directions based on our analysis.

\section*{Acknowledgment}
We thank the anonymous reviewers for their insightful
comments on our work. This work is supported by the Fundamental Research Funds for the Provincial Universities of Zhejiang (No.GK259909299001-004), the National Natural Science Foundation of China (Grants No.62402147, U2441239, U24A20336, 62172243, 62402425 and 62402418), the China Postdoctoral Science Foundation under No.2024M762829, the Zhejiang Provincial Natural Science Foundation under No.LD24F020002, the "Pioneer" and "Leading Goose" R\&D Program of Zhejiang, China (Grant No.2025C02261, 2025C02263, 2025C02033 and 2025C01082), the Zhejiang Provincial Priority-Funded Postdoctoral Research Project under No.ZJ2024001, and Zhejiang Provincial Key Laboratory for Sensitive Data Security Protection and Confidentiality Management No.2024E10048.



%\bibliographystyle{IEEEtran}
\bibliographystyle{IEEEtran}
\bibliography{sample-base}

\newpage
\appendices

\input{0x09Appendix}

\end{document}

%% file: 0x01Introduction.tex
\section{Introduction}
% \parasum{virtualization is prevalent}
Virtualization technology has been widely adopted in data centers, cloud computing, testing environments, and other scenarios~\cite{desai2013hypervisor}. These use cases place increasing demands on virtualization security, making it a critical focus for vendors and developers. As the cornerstone of virtualization, the hypervisor plays a pivotal role in managing and isolating virtual machines (VMs) within cloud environments, which makes it a high-value target for potential attackers. By exploiting vulnerabilities in the hypervisor, attackers can escalate privileges and gain control over the host system from a guest VM. This enables them to perform a range of malicious activities, such as data theft, system manipulation, or even a complete takeover of the cloud infrastructure—a scenario commonly referred to as \textit{virtual machine escape}. This form of attack has attracted growing attention from the security community, and numerous vulnerabilities and exploits have been uncovered in recent years~\cite{shao20203d, elhage2011virtunoid, zhao2019breaking, VMware-uhci, speedpwn, greatescapes, pan2021v, bulekov2024hyperpill, bulekov2022morphuzz}.

%[Previous] Among these vulnerabilities, \textit{pointer corruption} has emerged as a particularly dangerous and recurrent type. 
Among these vulnerabilities, \textit{pointer corruption} stands out as a particularly dangerous and common consequence, typically caused by issues like use-after-free (UAF) or out-of-bounds (OOB) memory access. These vulnerabilities can corrupt address values in memory, allowing attackers to hijack pointers and redirect them to attacker-controlled locations.
Such hijacked pointers may then be dereferenced to perform arbitrary memory access, enabling the attacker to read from or write to arbitrary locations in the hypervisor address space, potentially affecting sensitive data or control flow. Our investigation shows that pointer corruption is a frequent consequence of hypervisor vulnerabilities. An analysis of QEMU CVEs from the past five years indicates that about 23.9\% ultimately result in pointer corruption (see Table~\ref{tab:pointer-corruption}).

Despite their severity, exploiting pointer corruptions in hypervisors remains challenging due to indirection and complexity. Unlike in kernel or user-mode environments, exploitable memory structures in hypervisors are difficult to identify due to their scarcity and high degree of customization. To make matters worse, their address layout is further obscured by memory isolation mechanisms and address space layout randomization (ASLR), often requiring stringent conditions, such as additional vulnerabilities that leak address information, which are difficult to obtain. This renders existing exploitation frameworks, originally designed for kernel and user-mode environments~\cite{chen2020systematic,xie2025bridgerouter}, limited, as they typically rely on locating those exploitable structures. As a result, existing exploits targets hypervisor tend to be case-specific and handcrafted~\cite{a-ctf-style, speedpwn, greatescapes, VMware-uhci}, requiring deep manual analysis to identify suitable victim structures that expose read/write capabilities. To date, many of these vulnerabilities remain underexploited or entirely unexplored in practice, which limits our collective understanding and may cause security-critical bugs to be overlooked. This raises an alarming question: \textit{how to design a systematic framework for exploiting pointer corruption in hypervisor that generalizes across both diverse vulnerability types and hypervisor platforms?}

In this work, we conduct the first systematic study of exploiting hypervisor pointer corruptions through cross-domain memory interactions. Instead of identifying attack-controllable structure in host machines, we introduce guest memory as a reusable and attacker-controlled primitive. Our analysis reveals that the asymmetric isolation between host and guest in most modern virtualization systems can be leveraged to escalate exploitation. Specifically, while guest access to host memory is restricted, the host retains the ability to freely dereference pointers into guest memory. By redirecting corrupted pointers to crafted payloads in guest memory, attackers can reliably perform operations such as arbitrary memory writes, buffer manipulation, and fake object injection. We refer to this general exploitation paradigm applicable to most hypervisors as Cross-Domain Attacks (CDA). CDA transforms the challenge of finding exploitable host structures into a more tractable problem—leveraging guest-controlled memory as a stable, visible, and influenceable foothold for exploitation. Despite its power and generality, this attack surface has remained largely overlooked and lacks systematic exploration in prior work.

\begin{table}[ht]
\centering
\caption{Classification of QEMU CVEs (2019--2024) by root cause and corruption capability.}
\label{tab:pointer-corruption}
\resizebox{\linewidth}{!}{
\begin{tabular}{lcccc}
\toprule
\textbf{Vulnerability Category} & \textbf{Ptr. Corr.} & \textbf{Data-only Corr.} & \textbf{No Corr.} & \textbf{Total} \\
\midrule
Use-After-Free                  & 12 & 1  & 2  & 15 \\
OOB Write              & 18 & 12 & 0  & 30 \\
OOB Read               & 0  & 0  & 14 & 14 \\
Integer Overflow                & 1  & 5  & 1  & 7  \\
Uninitialized Variable          & 1  & 0  & 0  & 1  \\
Information Leak                & 0  & 0  & 13 & 13 \\
Logic/Crash and Others          & 0  & 0  & 54 & 54 \\
\midrule
\textbf{Subtotal}               & \textbf{32} & \textbf{18} & \textbf{84} & \textbf{134} \\
\textbf{Percentage}             & \textbf{23.9\%} & \textbf{13.4\%} & \textbf{62.7\%} & \textbf{100.0\%} \\
\bottomrule
\end{tabular}
}
\vspace{0.5em}
\begin{tablenotes}
\small
\item \textit{Ptr. Corr. =  vulnerabilities that may alter pointer values. Data-only Corr. =  vulnerabilities affecting data but not pointers. No Corr. = vulnerabilities with no corruptive effect on memory.}
\end{tablenotes}
\end{table}

To bridge the gap in understanding CDA in real-world, we develop a framework to automatically identify CDA-capable code paths and synthesize end-to-end exploits for different types of vulnerabilites that cause pointer corruptions. Starting from a PoC that triggers a corrupted pointer, our CDA framework locates hypervisor code segments that dereference guest-sourced pointers and identifies candidate gadgets for exploitation. It then applies a trace-guided input synthesis strategy to generate gadget-triggering inputs, and incrementally assembles complete exploits by aligning memory layouts and scheduling system interactions. Our system achieves a high degree of automation and demonstrates broad applicability across different hypervisor platforms and vulnerability types.

To evaluate the effectiveness and generality of CDA, we conducted comprehensive experiments. We first analyzed the distribution of cross-domain gadgets across multiple hypervisors and found that nearly all major attack surfaces expose code paths containing such gadgets, suggesting CDA’s wide applicability. We then measured the success rate of placing cross-domain gadgets in memory under different conditions. The results show that stack layouts exhibit substantial gadget coverage around common entry paths, while heap allocations provide even broader and more uniform opportunities for placing guest-derived pointers—making CDA-compatible layouts consistently achievable in practice. Additionally, we applied our technique to 15 real-world vulnerabilities—13 in QEMU and 2 in VirtualBox, and successfully achieved exploitation in all cases. These results confirm that CDA is broadly applicable and effective for exploiting hypervisor pointer corruption vulnerabilities. Finally, the automated components of CDA introduce only acceptable and predictable overhead.

\noindent {\bf The main contributions of this work are as follows:}

\begin{itemize}
    \item  A systematic characterization of cross-domain attacks (CDA), including their root causes and four pointer-use variants, establishing CDA as a general paradigm for upgrading pointer-corruption capabilities.
    
\end{itemize}

\begin{itemize}
    \item A prototype system is developed to automate the proposed exploitation approach. The experimental data are now available at an repository https://github.com/HDU-SEC/cda.
\end{itemize}

\begin{itemize}
    \item A comprehensive evaluation is conducted to demonstrate the coverage of cross-domain gadget identification within the hypervisor, the generality across various pointer corruption scenarios, and the effectiveness in exploiting real-world hypervisor vulnerabilities.
\end{itemize}

%% file: 0x02Background.tex
\section{Background}

\begin{figure}
    \centering
    \includegraphics[width=0.7\linewidth]{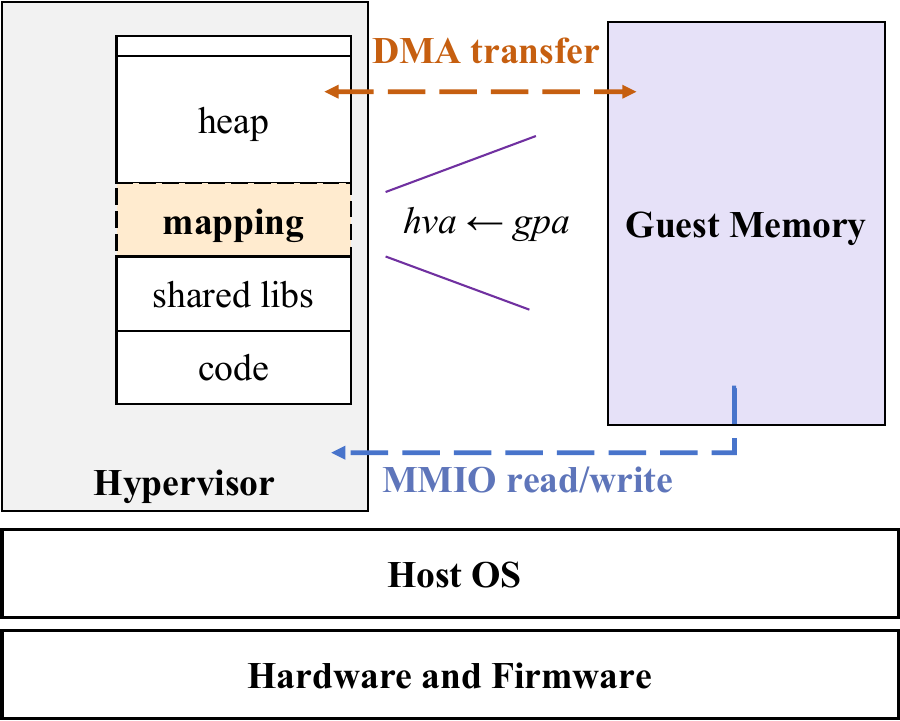}
    \caption{Architecture of Type-2 Hypervisors.}
    \label{fig:architecture}
\end{figure}

\subsection{Workflow of Hypervisor Exploitation} 
In virtualized environments, attackers typically exploit vulnerabilities in the hypervisor from a guest system in order to gain control over the host system, a process commonly referred to as \textit{virtual machine escape}. In this work, we focus on the typical Type-2 virtualization in the host, where the hypervisor runs as a user-space process and is thus subject to conventional user-space memory protection mechanisms, such as Address Space Layout Randomization (ASLR) and Non-Executable (NX). To exploit the hypervisor, attackers must construct a reliable exploitation chain that begins with a memory corruption vulnerability. This chain typically involves gradually upgrading capabilities, bypassing multiple layers of protection, and ultimately achieving arbitrary code execution within the hypervisor context.

\subsection{Guest Memory Management} 
Virtualization technology allows multiple virtual machines to share the same physical memory resources. The memory perceived by each guest operating system is actually an abstraction provided by the hypervisor through memory virtualization. As illustrated in Figure~\ref{fig:architecture}, the hypervisor typically allocates a contiguous region of memory in the host user space, referred to as the \textit{Host Virtual Address (HVA)} space, via \texttt{mmap}. This region is then exposed to the guest as its \textit{Guest Physical Address (GPA)} space. The \textit{GPA-to-HVA} mapping is registered with the host kernel via system calls, enabling the kernel to construct hardware-assisted page tables to support efficient memory address translation and access. From the host’s perspective, in addition to conventional user-mode memory segments, the hypervisor process includes a distinct memory-mapped region known as \textit{Guest Memory}, which is used to support the execution of the guest operating system and its applications. Through virtualization, the hypervisor enforces strict boundary isolation on this memory region: the guest can only access and manage its own guest memory, and is prohibited from accessing memory outside of its assigned space. This isolation ensures both memory safety and logical separation between the guest and the host. When communication between the guest and host is required, it is typically facilitated through Memory-Mapped I/O (MMIO) and Direct Memory Access (DMA). MMIO is used for issuing control commands, while DMA is employed for high-throughput data transfer.

\subsection{Exploiting Pointer Corruption Vulnerabilities} The basic concept of exploiting a pointer corruption vulnerability is to redirect a corrupted pointer to a memory region that enables the attacker to escalate their capabilities. Typically, the attacker targets a security-critical object, such as one that contains control information or a function pointer, and prepares the memory layout so that this object becomes accessible through the corrupted pointer. The pointer itself may be hijacked through direct overwrites, indirect memory manipulations, or side effects of other vulnerabilities. Once the corrupted pointer is dereferenced during subsequent execution, the attacker gains the ability to access or influence unintended memory regions, which can further lead to arbitrary reads or writes, unauthorized access, or control-flow manipulation.

\subsection{Scope and Assumption} 
\label{sec:assumption}

This work focuses on the automatic exploitation of pointer corruption vulnerabilities in type-2 hypervisors. We assume that the adversary has already obtained a pointer corruption primitive that allows modifying a host-side pointer via a guest-triggered vulnerability. The attacker is also assumed to possess full privileges within the guest VM, aligning with real-world cloud environments where attackers generally control the entire virtual machine. Additionally, we assume that standard user-mode protection mechanisms, such as address space layout randomization (ASLR) are enabled by default. The automatic capability upgrade technique proposed in this work does not violate these protections. Notably, type-2 hypervisor setting is widely used in multi-tenant cloud infrastructures (e.g., Alibaba Cloud~\cite{conver_image}) and aligns with prior hypervisor fuzzing research~\cite{bulekov2024hyperpill,ma2025truman}. Importantly, we do not assume that the attacker can directly execute arbitrary code on the host or invoke host-side functions at will. All exploitation steps must occur strictly along the natural logic of the hypervisor’s existing code paths, as triggered by legitimate guest–host interactions.

%% file: 0x03Motivation.tex
\section{Motivation and Our Exploitation}
In this section, we use a motivating example (§\ref{sec:example}) to motivate our exploitation (§\ref{sec:approach}) and clarify our assumptions (§\ref{sec:assumption}).

\subsection{A Running Example}
\label{sec:example}

\lstset{
  language=C,
  basicstyle=\ttfamily\small,
  keywordstyle=\bfseries\color{myblue},
  commentstyle=\itshape\color{gray},
  stringstyle=\color{orange},
  numbers=left,
  numberstyle=\tiny,
  numbersep=5pt,
  frame=single,
  breaklines=true,
  postbreak=\mbox{\textcolor{red}{$\hookrightarrow$}\space},
  showstringspaces=false,
  escapeinside={(*@}{@*)}, % for inline LaTeX commands
  morekeywords={size_t, uint8_t, uint16_t, void}
}

\begin{figure}[t]
\centering
\begin{lstlisting}
// guest-controlled offset
void usbredir_buffered_bulk_packet(..., uint8_t *data, size_t data_len, ...) {
    size_t i = choose_offset(...);
    ...
    bufp_alloc(dev, (*@\hl{data + i}@*), len, status, ep, data);  // interior ptr
}

int bufp_alloc(USBRedirDevice *dev, uint8_t *data, uint16_t len, ...) {
    ...
    if (bufpq_should_drop(dev, ep)) {
        (*@\hl{free(data);}@*)  // free(data + i): not chunk base
        return -1;
    }
    ...
}
\end{lstlisting}
\vspace{-0.5em}
\caption{Simplified illustration of the mistaken-free vulnerability in QEMU (CVE-2021-3682). 
An interior pointer \texttt{data+i} is passed downward and later freed, corrupting heap metadata.}
\label{fig:qemu_mistaken_free}
\end{figure}

Figure~\ref{fig:qemu_mistaken_free} presents a simplified example of the mistaken-free vulnerability (CVE-2021-3682) in QEMU. In this case, an interior pointer \texttt{data + i}, derived from a guest-controlled offset, is passed into the function \texttt{bufp\_alloc}. This value is received as the parameter \texttt{data}, which is later freed within the callee. As a result, the memory allocator incorrectly invokes \texttt{free()} on a non-base pointer, corrupting internal heap metadata. This behavior constitutes a classic case of \textit{pointer corruption}, and may be leveraged to manipulate subsequent allocations depending on allocator behavior and attacker-controlled buffer contents. At the time of writing, there is no public exploit available for this vulnerability.

Intuitively, to exploit this vulnerability, the attacker corrupts the internal metadata of a heap chunk, redirecting future allocations to a chosen address and causing them to overlap with critical in-memory structures. However, exploiting such pointer corruption in hypervisors is highly challenging. First, exploitable structures are rare and subject to strict constraints on chunk size and pointer offset, making reliable alignment difficult. Second, due to address space layout randomization (ASLR), the attacker lacks visibility into the runtime addresses of potential targets. Gaining precise control typically requires an additional information leak, significantly raising the bar for successful exploitation.

\textbf{Limitations of Existing Techniques}. The most related efforts on locating exploitable in-memory structures are found in kernel exploitation, where attackers have devised various strategies to identify and abuse kernel objects~\cite{chen2020systematic, xie2025bridgerouter}. However, such approaches are less applicable to hypervisors. The set of controllable in-memory structures is significantly smaller, and successful pointer hijacking often demands precise field alignment—constraints that vary across vulnerabilities and cannot be satisfied by generic layouts. These factors hinder the generalizability of structure-oriented search techniques in hypervisor contexts. To date, no systematic or automated exploitation methodology has been established for hypervisor environments.

\textbf{Overlooked Isolation Flaws.} However, existing exploitation techniques commonly assume strong guest-host memory isolation in the hypervisor, enforcing a strict separation between guest and host memory usage. In this model, attackers are restricted to user mode-like capabilities, significantly limiting the exploitation potential. Yet this assumption does not always hold in practice. Scavenger~\cite{pan2021scavenger} had already demonstrated that once a corrupted host pointer was redirected into guest memory, the hypervisor could access guest-controlled content without restriction. This broke isolation boundaries and allowed guest memory to participate in host-side heap operations, enabling far stronger primitives than previously considered.

\begin{tcolorbox}[
    colback=white,       
    colframe=gray!50,  
    coltitle=black,      
    boxrule=0.5pt,       
    arc=4pt,             
    left=6pt, right=6pt, top=4pt, bottom=4pt, 
    enhanced,
    sharp corners=south,  
    fontupper=\normalsize,
    drop shadow           
]
\textbf{New Insight:} Modern hypervisors exhibit a fundamental asymmetry: although a guest cannot directly read or write host memory, the host routinely accesses guest memory with few restrictions. This \textit{weak isolation} exposes a contiguous, attacker-controlled region directly within the host address space. From the attacker’s perspective, every byte of guest memory is controllable, making it a stable and highly manipulable foothold for exploitation. Redirecting a corrupted host pointer into this region circumvents the scarcity of suitably aligned host structures and sidesteps heap-layout uncertainty, opening a new avenue for precise and reliable hypervisor exploits.

\end{tcolorbox}

\subsection{Our Exploitation}
\label{sec:approach}

\begin{figure*}[!htbp]
    \centering
    \includegraphics[width=0.8\linewidth]{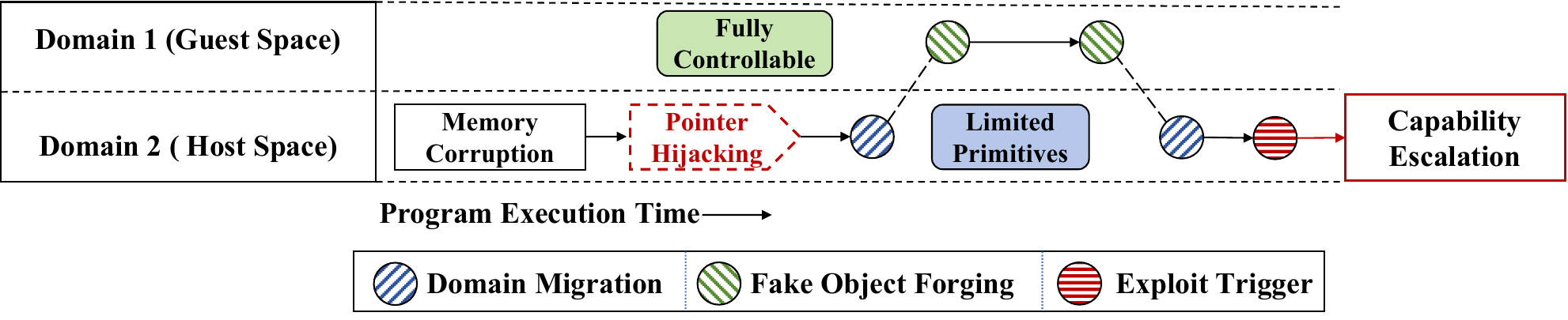}
    \caption{CDA approach for exploiting pointer hijacking primitives.}
    \label{fig:approach}
\end{figure*}

% --- 自定义颜色 ---
\definecolor{deepblue}{RGB}{0, 102, 204}
\definecolor{deepred}{RGB}{204, 0, 0}
\definecolor{codekey}{RGB}{0, 0, 180}
\definecolor{codecomment}{RGB}{100, 100, 100}
\definecolor{codestring}{RGB}{0, 150, 0}

% 背景色
\definecolor{guestbg}{RGB}{255, 245, 245} % 浅红 (Attacker)
\definecolor{hostbg}{RGB}{245, 250, 255}  % 浅蓝 (Host)

\lstdefinestyle{matchedstyle}{
    language=C,
    basicstyle=\ttfamily\small,
    commentstyle=\itshape\color{codecomment},
    numbers=none,          
    tabsize=4,
    showstringspaces=false,
    breaklines=true,
    frame=none,
    escapeinside={(*@}{@*)}, 
    keywords={void, int, if, return, else, while, struct},
    morekeywords=[2]{uint8_t, size\_t, uint16\_t, Req, MMIO\_ADDR}, 
    literate={_}{{\char`\_}}1, 
}

\begin{figure}[!t]
\begin{tcolorbox}[
    colback=white,          % 整体背景白
    colframe=gray!80!black, % 边框深灰
    coltitle=white,         % 标题文字白
    fonttitle=\large,
    enhanced,               % 启用高级绘图引擎
    drop shadow,            % 添加阴影
    sharp corners=south,    % 底部直角 (可选)
    rounded corners=north,  % 顶部圆角
    boxrule=0.5mm
]

    % === Guest (Attacker) 部分 ===
    \begin{tcolorbox}[
        colback=guestbg,    % 浅红背景
        colframe=guestbg,   % 隐藏内部边框
        left=2mm, right=2mm, top=0mm, bottom=0mm,
        arc=0mm, outer arc=0mm
    ]
        \textcolor{deepred}{\textbf{\sffamily // Guest (Attacker)}}
        \begin{lstlisting}[style=matchedstyle, numbers=none]
fake = guest_alloc_page();
    (*@\textcolor{deepred}{\ensuremath{\hookrightarrow}\ \textit{(1) allocate fake object}}@*)
mmio_write(MMIO_ADDR, map_gpa(fake));
    (*@\textcolor{deepred}{\ensuremath{\hookrightarrow}\ \textit{(2) send fake object's GPA}}@*)
...
fake->ops = final_attacker_ops;
    (*@\textcolor{deepred}{\ensuremath{\hookrightarrow}\ \textit{(5) modify fake object to finalize the exploit}}@*)
        \end{lstlisting}
    \end{tcolorbox}

    % === 分割线 ===
    \begin{center}
        \textcolor{gray!40}{\rule{0.9\linewidth}{0.5pt}} % 灰色细线
    \end{center}

    % === Host (Hypervisor) 部分 ===
    \begin{tcolorbox}[
        colback=hostbg,     % 浅蓝背景
        colframe=hostbg,    % 隐藏内部边框
        left=2mm, right=2mm, top=0mm, bottom=0mm,
        arc=0mm, outer arc=0mm
    ]
        \textcolor{deepblue}{\textbf{\sffamily // Host (Hypervisor)}}
        \begin{lstlisting}[style=matchedstyle, numbers=none]
s->ptr = (Req *)gpa_to_hva(gpa);
    (*@\textcolor{deepblue}{\ensuremath{\hookrightarrow}\ \textit{(3) vulnerability overwrites pointer to fake object}}@*)
qemu_free(s->ptr);
    (*@\textcolor{deepblue}{\ensuremath{\hookrightarrow}\ \textit{(4) host uses attacker object}}@*)
        \end{lstlisting}
    \end{tcolorbox}

\end{tcolorbox}

\caption{Code example to show the workflow of CDA.}
\label{fig:CDAexample_code}
\end{figure}

Driven by the new insights, we present a \textit{systematic characterization} of how pointer hijacking primitives in hypervisors can be exploited through Cross-Domain Attacks (CDA) — a class of techniques that enable capability escalation without relying on exploitable structures, crafted I/O primitives, or additional information leaks. It provides a more general and reusable exploitation strategy, significantly reducing the complexity of identifying usable structures and constructing fragile, highly specialized payloads, thus ensuring the success of complete exploitation.

Our approach involves two core steps: \textit{cross-domain migration} and \textit{fake object forging}, as illustrated in Figure~\ref{fig:approach}. In the first step, a pointer hijacking primitive is repurposed to redirect the host-side pointer to the guest memory space, enabling cross-boundary injection of attacker-controlled data. In the second step, this redirected pointer is used to access a carefully crafted fake object placed in guest memory, resulting in capability escalation.

\textbf{Step 1: Cross-Domain Migration.} It is achieved by placing a special type of code snippet that leaves a host virtual address (HVA) pointing to guest memory in the host address space. These snippets are part of the communication mechanism between the guest and the host, responsible for establishing the mapping between a GPA and a corresponding HVA, and for facilitating data transfer. We refer to such code snippets as \textit{cross-domain gadgets}. Through careful memory layout manipulation during the window between vulnerability triggering and pointer dereference, the residual guest-HVA pointer\footnote{A guest–HVA pointer is a host-side pointer whose value is a host virtual address (HVA) that corresponds to a guest memory region.} left by these gadgets can be reused to redirect the corrupted pointer into the guest memory space.

\textbf{Step 2: Fake Object Forging.} It is achieved by carefully placing a forged object within the guest memory, at the location pointed to by the hijacked pointer. This fake object mimics the layout and semantics expected by the hypervisor's internal logic—such as containing function pointers, capability flags, or structured data fields. When the corrupted pointer is later dereferenced by the host, the hypervisor transparently accesses this attacker-controlled payload, unknowingly operating on manipulated data. This enables the attacker to upgrade the exploit capability without modifying any host-side data or bypassing memory protections directly.

\textbf{Code Sample of CDA Workflow.}  As illustrated in the Figure~\ref{fig:CDAexample_code}, the guest first allocates a page to store a forged object and issues an MMIO write to trigger the vulnerable host code path (Steps 1–2). During this host-side processing, a memory corruption bug (e.g., overflow or UAF) allows the attacker to overwrite an internal hypervisor pointer with an arbitrary value. By supplying the host virtual address (HVA) corresponding to the forged guest page, the attacker forces the hypervisor to treat attacker-controlled guest memory as a legitimate host object (Step 3). When the hypervisor later performs an operation such as \textit{free()} or a structure dereference on this corrupted pointer (Step 4), it unknowingly operates on the forged object. The guest can then modify the same memory page at any time, enabling it to install the final payload that will be executed upon the hypervisor’s subsequent dereference (Step 5).

\begin{figure}
    \centering
    \includegraphics[width=\linewidth]{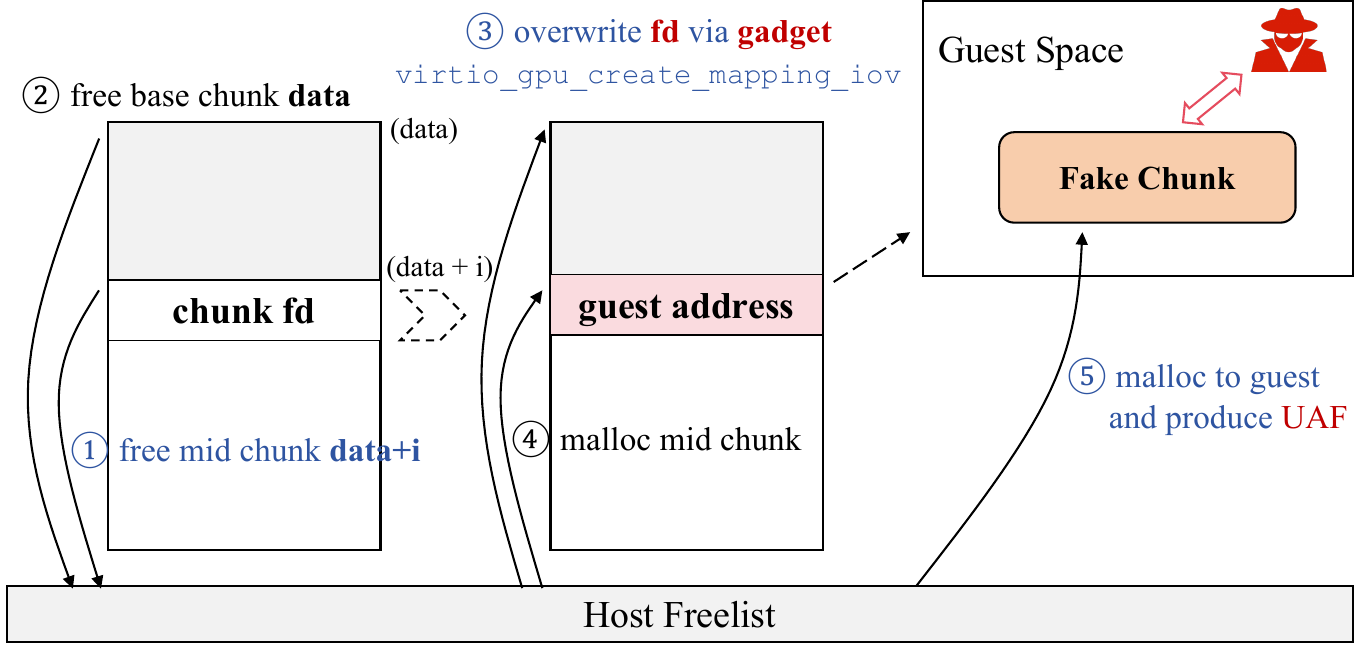}
    \caption{Exploiting the running example.}
    \label{fig:running-example}
\end{figure}

\textbf{Exploiting Running Example.} To exploit the vulnerability in the running example, we use the \texttt{virtio\_gpu\_create\_mapping\_iov} function as a cross-domain gadget, as illustrated in Figure~\ref{fig:running-example}. This gadget is capable of producing a guest-HVA pointer within a heap-allocated object, and its chunk size is compatible with the vulnerable structure.

When the vulnerability is triggered (\ding{172} in Figure~\ref{fig:running-example}), the attacker first frees a subregion within a heap buffer, specifically, the pointer at \texttt{data + i}, and then subsequently frees the original buffer \texttt{data}. This sequence creates overlapping chunks that simultaneously reside in the hypervisor’s freelist. Following this (\ding{174} in Figure~\ref{fig:running-example}), the attacker invokes the selected cross-domain gadget to allocate a new chunk and uses heap spraying to place it at the reclaimed data location. At this point, the metadata of the freed chunk at \texttt{data + i} is modified to overwrite its \texttt{fd} pointer with a guest address. This manipulation causes a fake chunk located in guest memory to be inserted into the host freelist. As a consequence (\ding{176} in Figure~\ref{fig:running-example}), the next heap allocation made by the hypervisor retrieves this fake chunk, effectively returning a pointer into guest-controlled memory. Since the attacker has full read/write control over guest memory, this transforms the vulnerability into a standard Use-After-Free primitive. The attacker can then reliably craft payloads within guest memory to corrupt critical host data structures or redirect control flow, thereby achieving a full virtual machine escape.

\textbf{CDA Variants.} To further demonstrate the expressiveness and versatility of the Cross-Domain Attack (CDA) model, we identify and categorize four representative variants, each characterized by the dereference semantics of the hijacked pointer and the resulting impact. These variants include arbitrary Code Execution (\textbf{CDA\textsuperscript{A}}), where the attacker gains control over the program counter or function pointer to execute arbitrary code; Information Leakage (\textbf{CDA\textsuperscript{I}}), which enables reading sensitive data by redirecting pointers to attacker-observable memory; Critical Data Overwriting (\textbf{CDA\textsuperscript{O}}), which targets security-critical fields such as credentials or control flags; and Chunk Confusion (\textbf{CDA\textsuperscript{C}}), which manipulates allocator metadata to corrupt heap integrity. These variants illustrate the broad range of exploitation consequences achievable through CDA under different vulnerability scenarios. A formal taxonomy and detailed case studies for each variant are provided in Appendix~\ref{sec:variants}.

\textbf{Technical Challenges.} To perform the exploitation described above, an adversary must address several key challenges. First, identifying valid gadget candidates is difficult due to their implicit and inconsistent patterns across the hypervisor codebase. Without explicit semantics or uniform signatures, systematically locating these translation points is highly non-trivial. Second, establishing a viable match between a corrupted pointer and a gadget requires understanding the memory characteristics of the gadget-produced pointer. Lacking such knowledge makes the matching process substantially more ambiguous and error-prone. Third, triggering deep gadget paths requires inputs that satisfy complex and hidden control-flow conditions. The vast input space and lack of intermediate feedback hinder direct exploration. Finally, integrating all components into a complete exploit demands precise coordination of control flow, memory layout, and time window.

%% file: 0x04Methodology.tex
\section{Methodology}

\begin{figure}
    \centering
    \includegraphics[width=\linewidth]{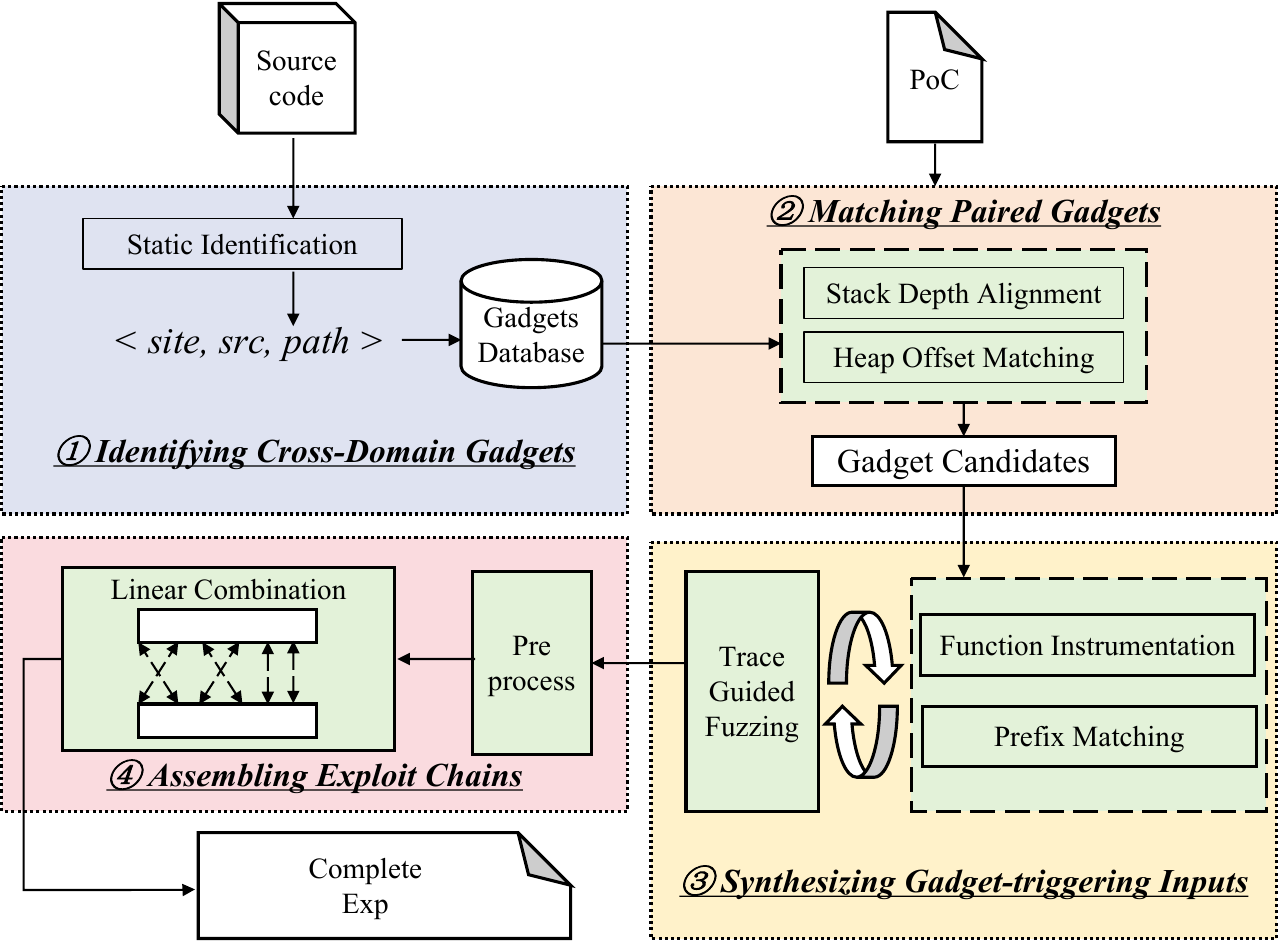}
    \caption{The overall workflow of CDA framework.}
    \label{fig:system-overview}
\end{figure}

Figure~\ref{fig:system-overview} outlines the workflow of the CDA framework, which is divided into four sequential stages. At a high level, the framework takes as input (i) the hypervisor’s source code and (ii) a proof-of-concept (PoC) that triggers a pointer corruption, and produces as output an exploit that redirects the corrupted pointer into the guest memory—achieving one of the four CDA variants. Starting from a pointer corruption vulnerability, we first statically identify cross-domain gadget instances within the hypervisor and organize them into a structured database (§\ref{sec:identifying-gadgets}). Given a PoC, we then match it with suitable gadget candidates based on memory layout compatibility—differentiating between heap- and stack-based corruption models (§\ref{sec:matching}). Next, we synthesize concrete guest inputs that reliably trigger the selected gadgets, using a trace-guided fuzzing strategy to incrementally drive execution through the intended call chains (§\ref{sec:synthesizing}). Finally, a complete exploit is assembled by integrating the original PoC and the gadget-triggering input, forming a coherent execution sequence that redirects the corrupted pointer to attacker-controlled guest memory and completes the capability escalation (§\ref{sec:assembling}). Subsequent operations after the pointer redirection are beyond the scope of this paper, as they follow standard exploitation procedures.

\subsection{Identifying Cross-Domain Gadgets}
\label{sec:identifying-gadgets}

Identifying cross-domain gadgets involves statically analyzing the hypervisor codebase to extract all code sites that perform GPA-to-HVA translation under guest influence. These translation functions map guest physical addresses to host virtual addresses, during which guest-HVA pointers are temporarily or globally stored in host memory. Empirically, we find that these sites represent the primary contexts where guest-derived addresses are likely to reside in host memory, creating opportunities for pointer redirection. In addition, we further identify their guest-accessible invocation paths. Cross-domain gadgets are thus necessary to bridge corrupted host pointers to attacker-controlled guest memory when ASLR prevents direct address targeting, capturing code-reuse scenarios where existing instructions leave guest addresses accessible. The full list of hypervisor functions responsible for address translation is provided in Appendix~\ref{sec:translation}.

\noindent\textbf{Definition: Cross-Domain Gadgets.} 
$\mathcal{G} = \{ (\mathit{site}, \mathit{src}, \mathit{path}) \mid \mathit{site} \in \mathcal{S}, \mathit{src} \in \mathcal{T}, \mathit{path} \in \mathcal{P} \}$ denotes the set of all cross-domain gadgets. Each element $g \in \mathcal{G}$ is a 3-tuple that uniquely identifies a gadget instance within the hypervisor. Here $\mathit{site}$ refers to a program location where a GPA-to-HVA translation is performed, $\mathit{src}$ denotes a guest-controllable operand that flows into the translation logic, and $\mathit{path}$ represents the static call chain from a guest-facing interface to the translation site.

To identify such gadgets in practice, we conduct a static analysis over the hypervisor codebase. We begin by locating all translation sites that perform GPA-to-HVA mappings, including standard functions and subsystem-specific logic. For each site, we trace the origin of the GPA operand to determine whether it can be influenced by the guest, either through descriptor structures, MMIO registers, or I/O buffers. Sites with no viable guest-controlled sources are discarded.

Focusing exclusively on \textit{guest-influenced translations} enables precise control over the resulting host virtual address. This controllability eliminates the need for imprecise memory spraying or blind guessing across the entire guest memory space, and allows the attacker to steer the corrupted pointer to a specific location in guest memory. As a result, this constraint significantly improves the reliability and precision of subsequent exploit stages.

For each remaining candidate, we extract its associated call chain using static source-to-sink analysis, which identifies guest-reachable paths that lead from external interfaces to the translation site. This path provides crucial context for evaluating gadget reachability and constructing triggering inputs in later stages. Each verified ($\mathit{site}$, $\mathit{src}$, $\mathit{path}$) triple is then inserted into a structured gadget database, forming the set $\mathcal{G}$.

\begin{table*}[t]
\centering
\caption{A cross-domain gadget extracted from QEMU's NVMe device.}
\label{tab:nvme_gadget_example}
\resizebox{\textwidth}{!}{
\begin{tabular}{@{}ccccccc@{}}
\toprule
\textbf{Gadget Family} &
\textbf{Upper Function} & 
\textbf{Translation Function} & 
\textbf{HVA Variable} & 
\textbf{GPA Source Field} & 
\textbf{Trigger Type} & 
\textbf{Call Path} \\
\midrule
DMA gadget &
dma\_memory\_write & 
address\_space\_write & 
ram\_ptr & 
s$\rightarrow$tx\_descriptor  & 
MMIO & 
\begin{tabular}[c]{@{}l@{}}
\texttt{nvme\_mmio\_write} $\rightarrow$ \texttt{nvme\_process\_db} \\
$\rightarrow$ \texttt{stl\_le\_pci\_dma} $\rightarrow$ \texttt{stl\_le\_dma} \\
$\rightarrow$ \texttt{dma\_memory\_write}
\end{tabular} \\
\bottomrule
\end{tabular}
}
\end{table*}

\textbf{Example.} Table~\ref{tab:nvme_gadget_example} presents a representative example of a cross-domain gadget extracted from QEMU’s NVMe device. The gadget is triggered via an MMIO write, where a guest-controlled field of type \texttt{dma\_addr\_t} is translated into a HVA and stored in a pointer variable named \texttt{ram\_ptr}. This pointer is later used as an access target and can be abused as a redirection target for a corrupted pointer during exploitation. Since \texttt{ram\_ptr} holds a guest-derived address, an attacker who hijacks a corrupted pointer to target \texttt{ram\_ptr} can effectively manipulate host logic to operate on attacker-controlled guest memory. This makes the gadget suitable in CDA-style exploitation. A full list of cross-domain gadgets are shown in Table~\ref{tab:cross_domain_gadgets} in the appendix.

\subsection{Matching Paired Gadgets}
\label{sec:matching}

Building upon the cross-domain gadget database, we now describe how to match a given pointer hijacking vulnerability with a suitable gadget instance to enable exploitation. The database contains numerous program paths that generate guest-address HVAs at different memory offsets, and the key insight is that our framework aims to \textit{pair a vulnerability with the gadget whose residual guest-HVA pointer is spatially aligned with the corrupted pointer}. This spatial alignment allows the gadget’s leftover HVA to overwrite the corrupted pointer exactly, thereby transforming an otherwise unexploitable corruption into a valid cross-domain redirection primitive.

Specifically, we focus on identifying a \textit{paired gadget} whose pointer write behavior can be aligned with the corrupted pointer produced by a proof-of-concept (PoC) input. Based on an empirical study of prior hypervisor vulnerabilities, we observe that corrupted pointers typically reside in either the stack or the heap. These two regions exhibit fundamentally different allocation semantics and memory layouts, which in turn necessitate distinct pairing strategies. For stack-resident pointers, layout compatibility depends on relative stack depth. For heap-resident pointers, it depends on object structure and field offset. We therefore design two complementary gadget matching procedures, one for each case, formalized in Algorithm~\ref{alg:match}.

\begin{algorithm}[t]
\caption{Paired Gadget Matching Strategy}
\label{alg:match}
\KwIn{Gadget database $\mathcal{G}$, corrupted pointer metadata $M$, pointer region $R \in \{\text{stack}, \text{heap}\}$}
\KwOut{Set of matched gadgets $\mathcal{G}_{\text{match}}$}

$\mathcal{G}_{\text{match}} \leftarrow \emptyset$ \;

\If{$R = \text{stack}$}{
    \ForEach{$g = (\mathit{site}, \mathit{src}, \mathit{path}) \in \mathcal{G}$}{
        $g.\mathit{depth} \leftarrow \text{Length}(\mathit{path})$ \;
    }
    $p_{\text{depth}} \leftarrow M.\text{stack\_depth}$ \;
    \ForEach{$g \in \mathcal{G}$}{
        \If{$g.\mathit{depth} = p_{\text{depth}}$}{
            $\mathcal{G}_{\text{match}} \leftarrow \mathcal{G}_{\text{match}} \cup \{g\}$ \;
        }
    }
}
\ElseIf{$R = \text{heap}$}{
    \ForEach{$g \in \mathcal{G}$}{
        \If{$\text{IsStoredAsStructField}(g)$}{
            $(g.\mathit{size}, g.\mathit{offset}) \leftarrow \text{ExtractStructInfo}(g)$ \;
        }
    }
    $(s_{\text{size}}, o_{\text{offset}}) \leftarrow M.\text{heap\_layout}$ \;
    \ForEach{$g \in \mathcal{G}$}{
        \If{$g.\mathit{size} = s_{\text{size}}$ \textbf{and} $g.\mathit{offset} = o_{\text{offset}}$}{
            $\mathcal{G}_{\text{match}} \leftarrow \mathcal{G}_{\text{match}} \cup \{g\}$ \;
        }
    }
}
\Return{$\mathcal{G}_{\text{match}}$}
\end{algorithm}

\textbf{Stack-Based Matching.}  
This strategy formalizes the intuition that if a gadget deposits a guest-HVA pointer at a specific stack depth, and the corrupted pointer resides at the same depth during the crash, then the two can be treated as aligned. That is, we assume that matching the relative call depth is sufficient to ensure that the memory locations coincide, thereby enabling layout reuse. For each gadget in the database, we first annotate its relative stack depth based on the length of its static call chain (lines 2–4). Given a crashing PoC, we extract the depth of the corrupted pointer within the call stack (line 5), and then select gadgets with matching stack depth (lines 6–8). These are returned as candidate paired gadgets that can be deployed via controlled call sequences.

\textbf{Heap-Based Matching.}  
In contrast to the stack, heap memory is governed by allocator behavior and object layout. Heap-based corruptions often target structure fields embedded within dynamically allocated chunks. Our insight is that if a gadget writes a guest-HVA pointer into a structure field with a known size and offset, and if a corrupted pointer resides at the same layout position, then the attacker can feasibly align the two via object spraying. We first identify gadgets that store guest-HVA pointers as part of structure fields by performing backward dataflow analysis (lines 9–11), annotating each with its structure size and offset. Then, from the PoC, we extract the corrupted object’s size and the field’s offset (line 12). Layout-aligned gadgets are selected as viable heap-matching candidates (lines 13–16).

\subsection{Synthesizing Gadget-triggering Inputs}
\label{sec:synthesizing}

\begin{algorithm}[t]
\caption{Trace-Guided Input Synthesis for Gadget Triggering}
\label{alg:trace_fuzz}
\KwIn{Gadget $G$ with call chain $[f_1, f_2, \dots, f_n]$}
\KwOut{Input $x$ that triggers $G$ via the intended call chain}

Instrument each function $f_i \in [f_1 \dots f_n]$ to update \texttt{bitmap[$i$]} on execution\;
Instrument $f_n$ with a termination signal (e.g., assertion/crash)\;
Initialize \texttt{corpus} $\gets \emptyset$\;

\While{not timeout}{
    $x \gets$ \textsc{MutateInput}()\;
    \textsc{RunGuest}($x$)\;
    $P \gets$ \textsc{LongestValidPrefix}($[f_1 \dots f_n]$, \texttt{bitmap})\;

    \If{$P$ is longer than any previous prefix}{
        Add $x$ to \texttt{corpus}\;
    }

    \If{$f_n$ reached via valid prefix}{
        \Return $x$\;
    }
}
\Return $\perp$ \tcp*{No trigger found within time bound}
\end{algorithm}

\begin{figure}
    \centering
    \includegraphics[width=\linewidth]{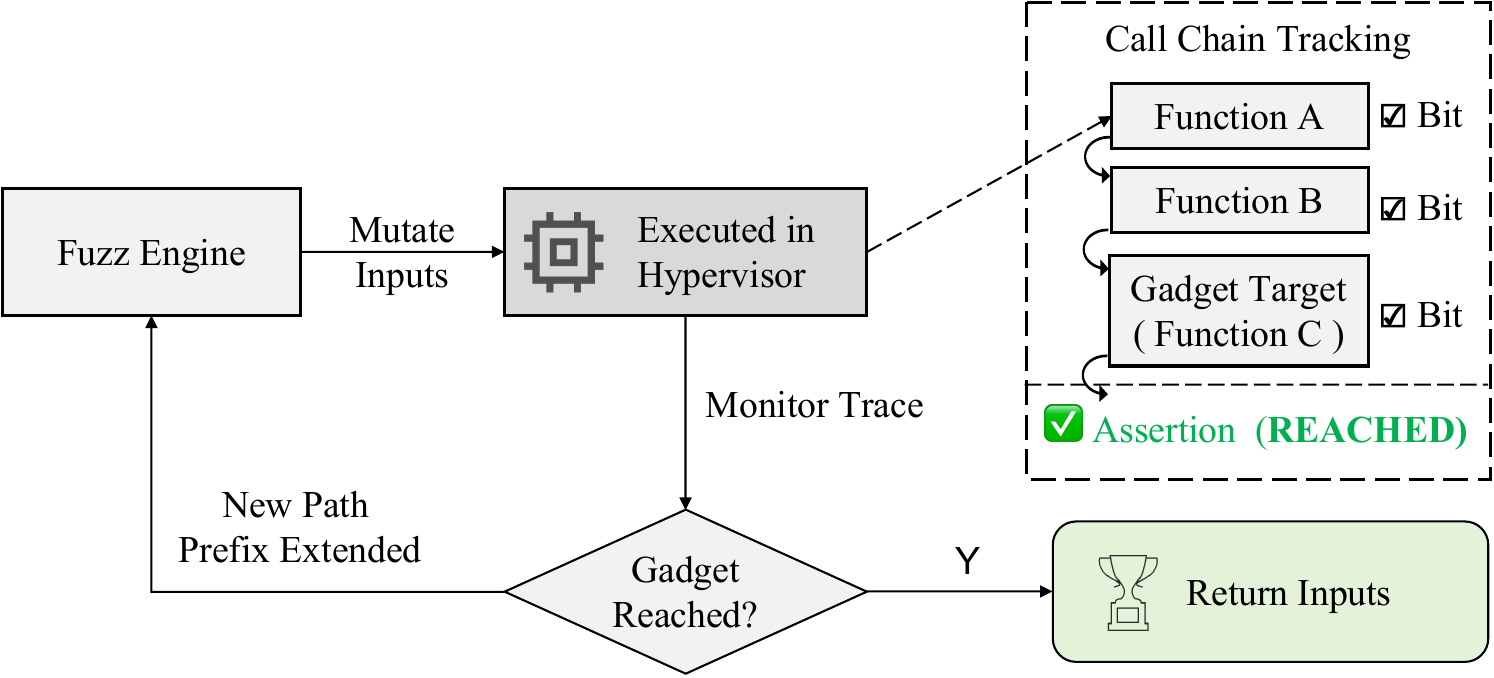}
    \caption{Trace-guided fuzzing workflow.}
    \label{fig:fuzzing}
\end{figure}

Given a set of matched gadget instances produced, the next step is to synthesize concrete guest inputs that can reliably trigger the execution of these gadgets. This is essential for enabling end-to-end exploit construction, but presents a practical challenge: gadgets typically reside deep within the hypervisor’s call chains, and the conditions under which they are executed may be highly constrained. Applying naive coverage-guided fuzzing is insufficient, as the search space is enormous and lacks effective guidance. To address this, we design a \textit{trace-guided input synthesis strategy} that incrementally steers fuzzing toward executing the target gadget, as shown in Algorithm~\ref{alg:trace_fuzz} and visualized in Figure~\ref{fig:fuzzing}. The core idea is that our strategy focuses on guiding the fuzzing process by monitoring progress along the call chain leading to the target gadget. Instead of blindly generating random inputs, the strategy tracks the path through the call chain and selectively preserves inputs that extend the chain towards the intended target. This turns the problem into a guided search over partial call chains, which significantly reduces the search space and improves the likelihood of successfully triggering the gadget.

The synthesis strategy consists of three key steps. \ding{172} For each candidate gadget, we instrument every function along its call chain to set a corresponding bit in a reserved region of the global coverage bitmap upon execution (line 1). The final target function—the gadget’s translation site—is instrumented with a termination signal \texttt{assert} to serve as a concrete indicator of successful triggering (line 2). \ding{173} During fuzzing, the system repeatedly generates and mutates guest inputs (line 5), then executes them within the virtualized environment (line 6), monitoring which chain elements have been traversed. After each execution, we extract the \textit{longest valid prefix} of the chain that has been traversed in the correct order (line 7). Inputs that extend this prefix are marked as \textit{interesting} and added to the fuzzing corpus (lines 8–9), promoting incremental exploration. This mechanism gradually drives the fuzzer toward deeper chain coverage. \ding{174} To ensure that the observed function visitation reflects the intended chain, we enforce strict caller verification: each function is only marked as reached if invoked directly by its expected predecessor. This constraint is embedded into the \nolinkurl{LongestValidPrefix} check (line 7) to eliminate spurious matches arising from unrelated execution paths. Once an input successfully reaches the gadget’s target function via the full intended chain (line 11), the fuzzer terminates and returns the corresponding input (line 12). The result of this stage is a collection of synthesized inputs corresponding to gadgets that have been confirmed executable through their annotated control paths.

% \textbf{Example.} xxx

\subsection{Assembling Exploit Chains}
\label{sec:assembling}

\begin{figure*}[!htbp]
    \centering
    \includegraphics[width=0.8\linewidth]{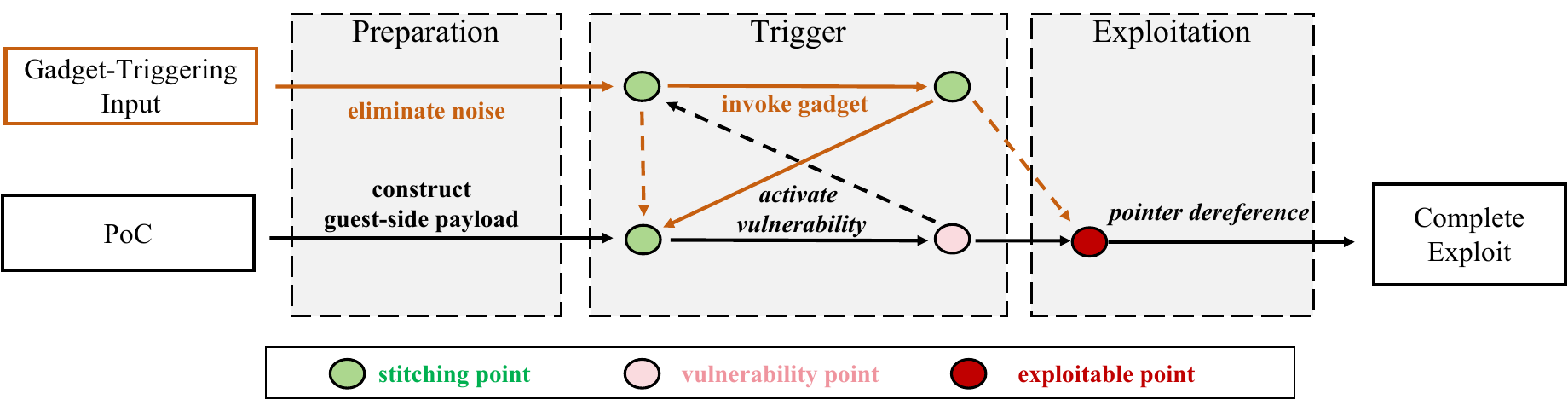}
    \caption{Workflow for assembling an exploit chain.}
    \label{fig:chaining}
\end{figure*}

Given a PoC that triggers a pointer corruption vulnerability and a synthesized gadget-triggering input, the final step is to construct an end-to-end exploit that redirects control or data flows across the guest–host boundary. This process involves stitching together the vulnerability trigger with a validated cross-domain gadget to form a coherent exploit chain.

We begin by decomposing the PoC into three semantic stages: the \textit{preparation phase}, which sets up memory layout and execution context; the \textit{trigger phase}, which produces the corrupted pointer; and the \textit{exploitation phase}, which dereferences the corrupted pointer. The preparation phase includes both payload placement and runtime stabilization. In heap-based scenarios, this entails spraying controlled structures into the guest heap to ensure consistent placement. In stack-based cases, we suppress asynchronous interference (e.g., signal delivery, multithread scheduling) by disabling preemption and isolating execution to a clean context, thereby ensuring a deterministic stack layout for the pointer corruption to manifest reliably.

To integrate the selected gadget, we insert the synthesized gadget-triggering input at a position in the PoC’s execution flow determined by the vulnerability type. The input is placed either before or after the corruption trigger, depending on the vulnerability type (e.g., post-trigger for UAF, pre-trigger for stack corruption). Our system accommodates both patterns by supporting flexible interleaving of the gadget-triggering input with the PoC logic. This ensures that the corrupted pointer, once created, is redirected to a valid and attacker-controlled guest memory region by invoking a gadget instance identified in \ref{sec:identifying-gadgets} and matched in \ref{sec:matching}. This intermediate step preserves execution continuity while re-routing the pointer dereference target, effectively bridging the gap between vulnerability and capability elevation.

Following gadget invocation, the program proceeds to dereference the now-translated pointer. Depending on the semantics of the gadget and the nature of the dereference, this access may directly manipulate sensitive data structures or initiate privilege escalation. The overall exploit chain, which includes vulnerability activation, gadget activation, and capability upgradation, is illustrated in Figure~\ref{fig:chaining}, which shows the temporal and logical structure of this composition.

\subsection{Extensibility and Portability}

Our framework is designed with modularity and adaptability in mind, enabling it to generalize across different hypervisors and vulnerability types. To port the framework to a new hypervisor, developers only need to perform lightweight and localized adaptations while keeping the core pipeline for gadget identification, matching, and exploit synthesis unchanged. Specifically, our static analysis step requires extracting three categories of information from the hypervisor’s developer documentation: (1) the \textbf{base translation functions} (e.g., \texttt{bhyve}’s \texttt{vm\_map\_memory}), (2) the \textbf{I/O entry points} (e.g., \texttt{e1000\_write\_reg}), and (3) the \textbf{types and definitions of guest-influenced variables} (e.g., \texttt{vm\_paddr\_t}).

%% file: 0x05Evaluation.tex
\section{Evaluation}
To demonstrate the practicality of CDA, we collected a series of metrics to establish its prevalence in virtualization environments. We focused on QEMU (used either standalone or together with KVM) and VirtualBox, which are representative hypervisor software. Our goal is to address the following research questions:

\begin{itemize}
\item \textbf{RQ1:} How prevalent are cross-domain gadgets, and what is their distribution within hypervisors?

\item \textbf{RQ2:} How frequently do cross-domain gadgets produce guest–HVA pointers?

\item \textbf{RQ3:} How does CDA perform in actual exploit scenarios?
\end{itemize}

For our static analysis, we utilized CodeQL, while function stack frames and variable offsets were extracted at the binary level using Ghidra. To validate the actual exploit scenarios, each experiment was conducted on a server equipped with an Intel(R) Xeon(R) Platinum 8124M CPU @ 3.00GHz (72 cores) and 64GB of RAM. Both the host and guest operating systems were 64-bit Ubuntu 20.04 LTS.

\begin{figure}
    \centering
    \begin{subfigure}[b]{0.22\textwidth}
        \includegraphics[width=\linewidth]{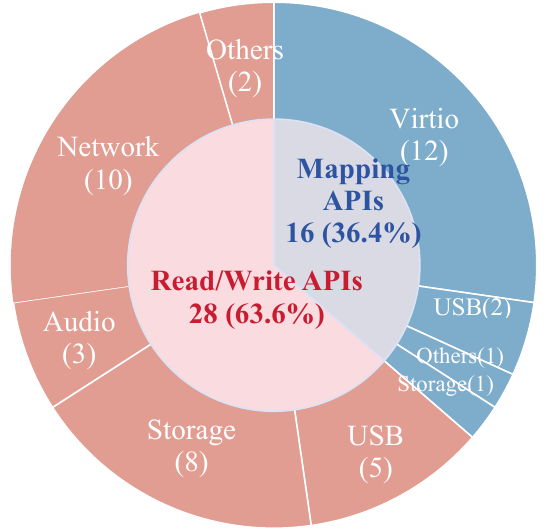}
        \caption{Distribution in QEMU.}
        \label{fig:qemu-distribution}
    \end{subfigure}
    \hspace{2mm}
    \begin{subfigure}[b]{0.22\textwidth}
        \includegraphics[width=\linewidth]{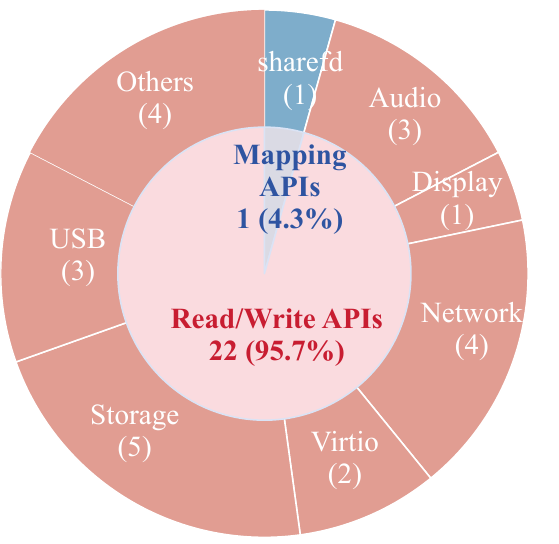}
        \caption{Distribution in VirtualBox.}
        \label{fig:virtualbox-distribution}
    \end{subfigure}
    \caption{Statistics on the number of virtual devices with GPA-to-HVA translation functions in the hypervisor.}
    \label{fig:distribution}
\end{figure}

\subsection{Gadget Prevalence}
\label{sec:prevalence}

To understand how cross-domain gadgets originate inside real hypervisors, we begin with the static analysis stage of our CDA framework. First, we extract all GPA-to-HVA translation functions in both QEMU and VirtualBox. These are all APIs that accept guest-supplied GPAs and convert them into host-accessible addresses under guest influence. Using documentation-guided patterns and interprocedural tracing, our analysis identifies \textbf{44 translation functions in QEMU} and \textbf{23 in VirtualBox}. Given that virtual devices represent the primary attack surface and constitute the largest portion of the hypervisor’s code~\cite{pan2021v,liu2023videzzo,bulekov2024hyperpill}, we use device type as the basic unit of analysis when categorizing these translation functions. Across both systems, these translation functions fall empirically into two major semantic categories: \textbf{read/write APIs} and \textbf{mapping APIs}.

\lstset{
  language=C,
  basicstyle=\ttfamily\small,
  keywordstyle=\bfseries\color{myblue},
  commentstyle=\itshape\color{gray},
  stringstyle=\color{orange},
  numbers=left,
  numberstyle=\tiny,
  numbersep=5pt,
  frame=single,
  breaklines=true,
  postbreak=\mbox{\textcolor{red}{$\hookrightarrow$}\space},
  showstringspaces=false,
  escapeinside={(*@}{@*)},
  morekeywords={size_t, uint8_t, uint16_t, void, dma_addr_t}
}

\begin{figure}[t]
\centering
\begin{lstlisting}
int lsi_mem_read(LSIState *s, dma_addr_t addr, void* buf, dma_addr_t len)
{
    if (s->mode & LSI_DMODE_SIOM) {
        (*@\hl{address\_space\_read(*s->pci\_io\_as, addr, buf, len);}@*)
    } else {
        (*@\hl{pci\_dma\_read(s, addr, buf, len);}@*)
    }
}
\end{lstlisting}
\vspace{-0.5em}
\caption{A gadget in QEMU's \texttt{lsi\_mem\_read} function that conditionally invokes either \texttt{address\_space\_read} or \texttt{pci\_dma\_read} based on the DMA mode. }
\label{fig:stack-case}
\end{figure}

As shown in Figure~\ref{fig:distribution}, \textbf{read/write APIs constitute the majority—63.6\% in QEMU and 95.7\% in VirtualBox}. These APIs appear broadly across device families such as storage, USB, network, and block devices. Their high frequency reflects their fundamental role in moving data across the guest–host boundary. Because these APIs are embedded in many device I/O paths, they create numerous translation points through which guest-controlled GPAs flow. For example, Figure~\ref{fig:stack-case} presents code from the \texttt{lsi} storage device in QEMU, where calls to \texttt{pci\_dma\_read} and \texttt{address\_space\_read} temporarily materialize guest addresses as local variables during execution—an inherent consequence of C’s calling conventions.

\textbf{Mapping APIs account for the remaining 36.4\% of translation functions in QEMU and 4.3\% in VirtualBox.} Although fewer in number, mapping APIs often generate structured heap-allocated objects, such as descriptor tables or DMA mapping entries, in which guest addresses are embedded. These objects persist beyond the scope of individual API calls and may be reused or accessed by multiple devices, giving mapping-derived translation points a broader and more durable influence. In QEMU, mapping APIs are primarily used in virtio and USB subsystems, consistent with the design of high-throughput and paravirtualized devices. VirtualBox, by contrast, uses mapping APIs almost exclusively in the HGCM shared-file subsystem, reflecting its more centralized architecture. While our analysis also observes mapping patterns in initialization-only structures (e.g., \texttt{deviceState}), these memory regions are not attacker-controlled and are therefore excluded from our statistics.

Second, using these translation sites as anchors, we enumerate all concrete cross-domain gadgets on QEMU. For each translation function, we trace the propagation of guest-influenced GPAs along its call chain and record the contexts where these translated values interact with host pointer operations. This process yields \textbf{772 gadget instances}, which naturally cluster into \textbf{eight major gadget families}, each representing a distinct translation semantic and operand-usage pattern. Table~\ref{tab:cross_domain_gadgets} in the appendix summarizes these families, their representative APIs, and the underlying translation semantics.

Overall, this analysis shows that GPA-to-HVA translation logic is deeply embedded across hypervisor device-emulation code, and both read/write and mapping APIs serve as systematic and recurring sources of cross-domain gadgets. The large number and diversity of these gadgets highlight the structural ease with which guest-controlled addresses permeate hypervisor execution paths, underscoring the practicality of constructing CDA attacks across a wide range of devices and translation contexts.

\subsection{Pointer Presence}

Since our CDA exploitation pipeline relies on reusing guest-derived pointers left in the hypervisor’s memory, the availability and distribution of such pointers fundamentally determine whether a corrupted host pointer can be redirected to guest memory. To assess the feasibility of CDA across real virtualization workloads, we perform a comprehensive measurement of guest-pointer coverage in QEMU’s stack and heap.

\begin{figure}
    \centering
    \begin{subfigure}[b]{0.5\textwidth}
        \includegraphics[width=\linewidth]{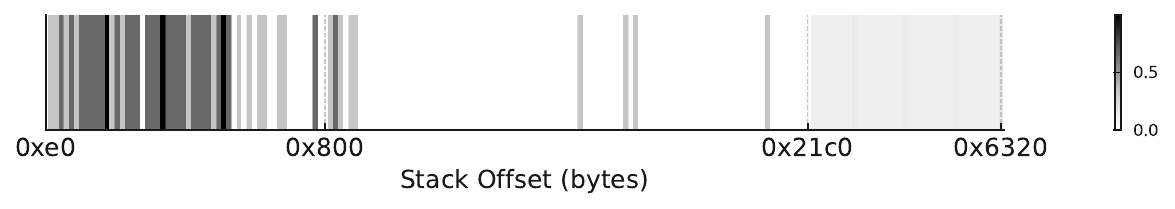}
        \caption{MMIO entry}
        \label{fig:stack-mmio-coverage}
    \end{subfigure}
  
    \begin{subfigure}[b]{0.5\textwidth}
        \includegraphics[width=\linewidth]{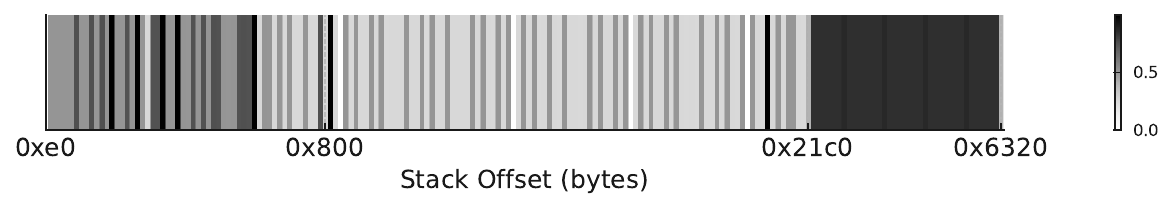}
        \caption{Timer/BH entry.}
        \label{fig:stack-bh-coverage}
    \end{subfigure}
    \caption{Distribution and coverage of guest-derived pointers in the QEMU stack.}
    \label{fig:stack-coverage}
\end{figure}

\begin{figure}
    \centering
    \includegraphics[width=\linewidth]{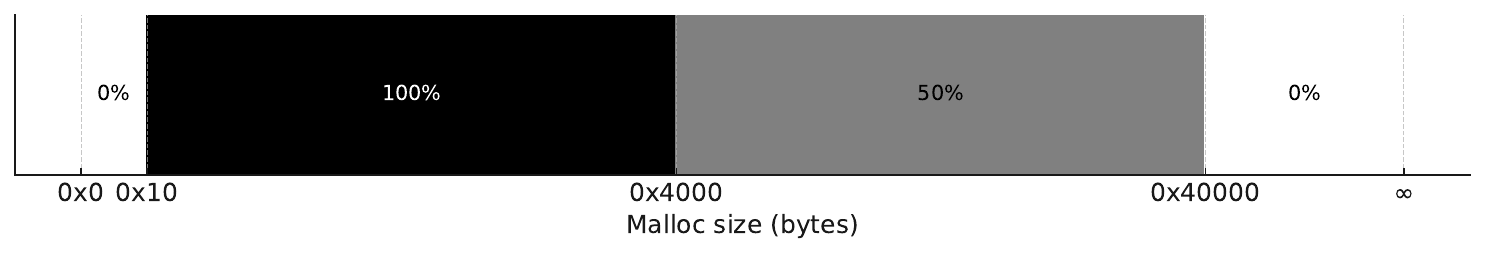}
    \caption{Heap coverage of guest-derived pointers in QEMU.}
    \label{fig:heap-coverage}
\end{figure}

\subsubsection{Stack Analysis}
To evaluate where guest-derived pointers naturally appear during hypervisor execution, we analyzed the stacks of QEMU’s three major entry paths—\texttt{MMIO} (memory-mapped I/O handlers), \texttt{bh} (bottom-half handlers), and \texttt{timer} (timer callbacks)—and aggregate all guest-pointer residues beneath their entry frames. Because \texttt{MMIO} uses a dedicated VCPU-thread stack while \texttt{bh} and \texttt{timer} callbacks share the main-loop stack, we analyzed them separately. \texttt{MMIO} handlers are invoked on VM exits, whereas \texttt{bh/timer} callbacks execute asynchronously on the main-loop thread. The resulting heat-strip visualizations (Figure \ref{fig:stack-coverage}) reveal a consistent and structured distribution pattern.

Across all entry paths, \textbf{shallow stack frames (roughly 0–0x200)} show relatively sparse residual pointers for \texttt{MMIO} due to multiple initialization layers, whereas \texttt{bh/timer}—invoked directly from the event loop—exhibit noticeably denser activity in the same region. \textbf{Mid-stack frames (approximately 0x200–0x800)} form a clear high-density band for all entry types, corresponding to the core device-emulation and DMA read/write routines that frequently spill guest-HVA pointers. Beyond this point, the \textbf{0x800–0x21c0} interval appears as a lighter, more heterogeneous transition region. This segment contains fewer residues and reflects device-specific behaviors; nonetheless, residues still appear intermittently, providing occasional placement candidates for CDA. A markedly different pattern emerges in the \textbf{deep stack region (around 0x21c0–0x6320)}. As shown in the figure~\ref{fig:stack-coverage}, this band extends even further under \texttt{bh/timer} execution, where virtio backends produce a long, continuous dark band driven by regularly structured DMA and DMA-mapping APIs, which leave guest addresses at both aligned and non-aligned offsets. This deep region is not only broader but also more uniform for \texttt{bh/timer} than \texttt{MMIO}, creating a large space where corrupted host pointers can reliably be redirected into guest memory.

Overall, these stacked dense bands—spanning shallow, mid, transition, and deep regions—show that QEMU repeatedly leaves guest-derived pointers across large portions of its execution stacks. Consequently, CDA remains feasible for pointer corruptions occurring at a wide variety of stack depths, regardless of the specific device or entry path involved.

\subsubsection{Heap Analysis}

We further examine where guest-derived pointers appear in QEMU’s heap allocations. By grouping allocations by size and visualizing pointer residues (Figure~\ref{fig:heap-coverage}), we observe a clear and highly structured pattern of coverage across malloc size classes. Allocations in the \textbf{0x10–0x4000} range exhibit \textbf{consistently full coverage}, meaning guest-controlled pointers can be positioned reliably at these sizes. In the \textbf{0x4000–0x40000} range, coverage remains substantial—typically around half of all possible positions—due to regular spacing and layout properties of QEMU’s allocation structures. Beyond these ranges, residues become scarce, and coverage drops to zero.

Our measurement is based directly on malloc size classes, reflecting how real heap manipulation works. In practice, objects of the same allocation size can be reliably steered into the same bins and co-located via standard heap fengshui techniques. This allows an attacker to position a corrupted heap object onto a chunk that already contains guest-pointer residues, making matching feasible whenever the allocation sizes align. Thus, evaluating coverage at the granularity of malloc size directly indicates which heap objects can practically host residue-bearing chunks during exploitation.

A key observation is that a wide variety of virtio backends naturally create \textbf{elastic guest-pointer placement opportunities} in these size classes: their internal structures and padding patterns allow the guest to control both the size and alignment of embedded addresses, resulting in a broad, contiguous spectrum of heap objects in which guest-HVA pointers can be positioned. Importantly, these patterns are not tied to a single data structure or device type; instead, they emerge repeatedly across multiple virtio implementations, making the effect systematic rather than device-specific. A representative example illustrating this elastic layout behavior is provided in Appendix~\ref{sec:elastic}.

Since most heap objects involved in I/O processing fall within these size ranges, nearly all heap-based pointer corruptions in QEMU can be paired with suitable guest-pointer residues. Consistent with this observation, our automated search identifies over a hundred residuable heap locations within one hour, confirming that heap layouts provide extensive and repeatable opportunities for CDA redirection.

\begin{table*}[]
\caption{We successfully exploited 15 real-world vulnerabilities using CDA, with 13 of them being in QEMU and 2 in VirtualBox. Among them, we use \textbf{A} to represent executing weird machine (Fig \ref{fig:variant1}), \textbf{I} to represent information leakage (Fig \ref{fig:variant2}), \textbf{O} to represent overwriting critical data (Fig \ref{fig:variant3}), and \textbf{C} to represent chunk confusion (Fig \ref{fig:variant4}). These are the types of CDA variants used for these vulnerabilities.}
\label{tab:exploitation}
\centering
\resizebox{0.75\linewidth}{!}{
\begin{tabular}{ccccccc}
\toprule
                            & \textbf{CVE id/Name} & \textbf{Device}       & \textbf{Vulnerability Type} & \textbf{CDA Variants} & \textbf{Impact}    & \textbf{Success}             \\
\midrule
\multirow{15}{*}{QEMU}      & CVE-2019-6778        & slirp                 & Heap overflow               & \textbf{O},\textbf{C}                  & RCE      & \checkmark           \\
                            & CVE-2019-14378       & slirp                 & Heap overflow               & \textbf{O},\textbf{C}                   & RCE     & \checkmark        \\
                            & CVE-2020-7039        & slirp                 & Heap overflow               & \textbf{O},\textbf{C}                   & RCE     & \checkmark          \\
                            & CVE-2020-14364       & USB                   & OOB                         & \textbf{A},\textbf{I}                  & RCE     & \checkmark         \\
                            & CVE-2021-3682        & USB redirector device & Mistake free                & \textbf{O},\textbf{C}                  & RCE      & \checkmark     \\
                            & CVE-2021-3929        & Nvme                  & UAF                         & \textbf{O},\textbf{C}                  & RCE      & \checkmark        \\
                            & CVE-2023-3180        & virtio-crypto         & Heap   overflow             & \textbf{A},\textbf{I}                   & RCE     & \checkmark        \\
                            & CVE-2023-6693        & virtio-net            & Stack overflow              & \textbf{I}                     & Info leak        & \checkmark      \\
                            & Scavenger            & NVMe                  & Uninitialized free          & \textbf{O},\textbf{C}                  & RCE     & \checkmark       \\
                            & Fixes: 1733eebb9e7   & virtio-iommu          & OOB read                   & \textbf{I}                    & Info leak          & \checkmark            \\
                            & CVE-2024-3446        & virtio-gpu            & Double   free               & \textbf{O},\textbf{C}                  & RCE     & \checkmark       \\
                            & CVE-2024-8612            & virtio-blk            & OOB read                 & \textbf{I}                     & Info leak     & \checkmark          \\
                            & Fixes: 62dbe54c         & virtio-sound          & Heap overflow               & \textbf{A}                     & RCE         & \checkmark            \\
\midrule
\multirow{2}{*}{VirtualBox} & CVE-2020-2575        & usb-ohci              & Uninitialized   heap        & \textbf{A}                     & RCE            & \checkmark                   \\
                            & CVE-2020-2758        & VHWA                  & UAF                         & \textbf{A},\textbf{I}                 & RCE     & \checkmark              \\
\bottomrule
\end{tabular}
}
\end{table*}

\subsection{Exploit Practicality}

To comprehensively evaluate the practicality of our approach, we examined all pointer-corruption vulnerabilities listed in Table~\ref{tab:pointer-corruption}, and successfully exploited 15 vulnerabilities in QEMU and VirtualBox. Our framework successfully exercised all four CDA variants across a broad spectrum of vulnerability classes, including heap overflows, out-of-bounds accesses, UAF, double free, and uninitialized free, and across diverse device types such as slirp, USB, NVMe, and multiple virtio subsystems. Among the 15 evaluated vulnerabilities in QEMU and VirtualBox, CDA enabled reliable exploitation in all cases, including several vulnerabilities that had previously been considered highly challenging to exploit in practice, such as CVE-2021-3682, CVE-2020-2575, and Scavenger. These results demonstrate that guest memory can serve as a stable and reusable primitive when traditional host-centric strategies fail. The CDA variant used for each vulnerability is summarized in Table~\ref{tab:exploitation}.

A closer examination of the variant distribution reveals clear patterns that align with the structural characteristics of different vulnerability types. CDA\textsuperscript{I} most frequently appears in out-of-bounds read vulnerabilities because redirecting a corrupted pointer to guest memory naturally produces attacker-controlled information leaks. CDA\textsuperscript{O} is widely applicable to overwrite-based bugs, where pointer-adjacent fields such as lengths, flags, or buffer descriptors can be reliably redirected to guest memory. CDA\textsuperscript{C} often arises in UAF-style vulnerabilities, including double free and uninitialized free, since the attacker can steer guest-mapped chunks into the host’s freelist and effectively convert guest objects into host-allocated ones. CDA\textsuperscript{A} appears less frequently because control-sensitive function pointers are relatively sparse in hypervisor structures; however, when such pointers exist, CDA\textsuperscript{A} enables reliable control-flow hijacking, as demonstrated in CVE-2020-2575 and CVE-2023-3180. These observations indicate that CDA variants occur naturally based on the underlying bug structure, and CDA provides a unified exploitation strategy that adapts consistently across them.

\begin{figure}
    \centering
    \includegraphics[width=0.65\linewidth]{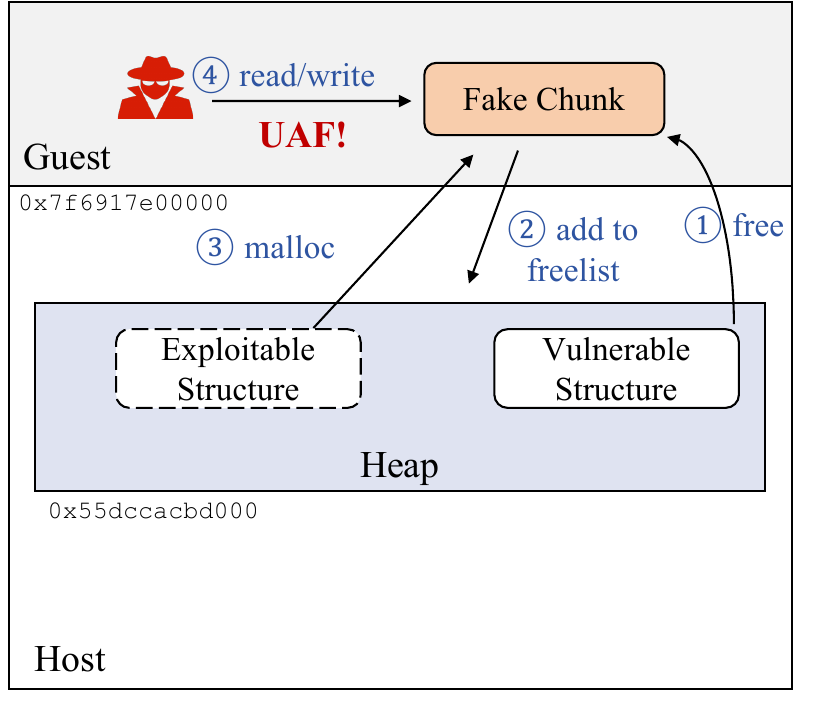}
    \caption{The workflow of QEMU NVMe case's exploitation based on the CDA attack.}
    \label{fig:exploit-case}
\end{figure}

\textbf{Case Study: QEMU NVMe Uninitialized Free.} The bug occurs when an \texttt{sglist} structure is freed without proper initialization, causing the hypervisor to free an attacker-influenced pointer. Under conventional exploitation models, the attacker would need to find a host-side structure that matches the size and layout of \texttt{sglist}, place it at an appropriate heap location, and identify device-specific primitives capable of manipulating its pointer fields. These conditions are difficult to satisfy in practice. CDA removes these obstacles. As illustrated in Figure~\ref{fig:exploit-case}, our framework identifies a virtio-based cross-domain gadget that leaves guest addresses in allocator-sized chunks, representing the exploitable structures. Through standard heap grooming, this gadget output is placed into the host’s freelist, ensuring that the vulnerable \texttt{sglist} structure receives a guest-derived pointer when allocated. During the uninitialized free, this guest-controlled fake chunk is inserted into the host’s freelist and becomes indistinguishable from a regular host allocation, giving the attacker full read-write access to a chunk now treated as host memory.

This transformation captures the core capability of CDA\textsuperscript{C}. Once the uninitialized free is redirected to a guest-controlled chunk, the vulnerability becomes a stable UAF primitive that already provides full attacker control over the freed object. From this point onward, the remaining exploitation steps follow well-established and standardized UAF exploitation workflows, such as steering the reallocated chunk toward security-critical structures or converting the primitive into control-flow hijacking. When stronger capabilities are desired, this UAF condition can also be transformed into other CDA variants. CDA\textsuperscript{O} enables metadata manipulation, CDA\textsuperscript{I} enables targeted information disclosure, and CDA\textsuperscript{A} becomes possible when a callable field is present. The NVMe case therefore illustrates how CDA turns a structurally difficult vulnerability into a reliable exploitation substrate, while still allowing further capability escalation through other CDA variants when needed.

\textbf{Analysis of Failed Cases.} During our evaluation, we also analyzed the remaining cases in Table~\ref{tab:pointer-corruption} where CDA was not achieved. We manually examined all the listed pointer corruption vulnerabilities and found that the failures generally fall into two categories: (1) some corrupted pointers are never dereferenced afterward, preventing further exploitation; (2) in other cases, the corrupted values are not attacker-controllable, making it infeasible to redirect it to a usable payload.

\begin{table}[t]
\centering
\small
\begin{tabular}{lccc}
\toprule
\textbf{CVE id/Name} & \textbf{Device} & \textbf{Input Gen.} & \textbf{Exploit Build}  \\
\midrule
CVE-2024-3446  & virtio-gpu & 268 s  & 17.8 min \\
CVE-2024-8612  & virtio-blk & 274 s  & 18.3 min \\
CVE-2021-3682  & USB-redir& 517 s  & 19.2 min \\
Scavenger      & NVMe & 531 s  & 14.6 min \\
CVE-2023-3180  & virtio-crypto & 522 s  & 18.9 min \\
\bottomrule
\end{tabular}
\caption{Dynamic runtime (10 runs) of input synthesizing and exploit assemblying across representative vulnerabilities.}
\label{tab:dyn-cost}
\end{table}

\subsection{Overhead Analysis of the CDA}

We measured the dynamic overhead of CDA by evaluating two steps: input synthesizing (Sec.\ref{sec:synthesizing}) and exploit assembling (Sec.\ref{sec:assembling}), as shown in~\ref{tab:dyn-cost}. Five representative vulnerabilities across QEMU and VirtualBox were selected to cover typical device types and bug categories. For input generation, we report the average time (ten runs) required to find the first usable, low-noise cross-domain gadget. The results show stable completion times ranging from 268 to 531 seconds, depending on gadget-family complexity. For exploit building, the automated combination of primitives incurs negligible cost. However, a small manual adjustment is required to realign heap-grooming offsets, since certain gadget-triggering inputs introduce minor side effects that slightly shift the heap layout. This validation step results in an overall average build time of 14.6 to 19.2 minutes across vulnerabilities. Overall, these measurements show that CDA incurs an acceptable and predictable dynamic overhead, enabling end-to-end exploit construction with manageable cost.

%% file: 0x06Discussion.tex
\section{Discussion}

\subsection{Possible Defense Mechanism}

\textbf{Memory Access Control}. We believe that preventing the host from accessing guest memory by default could serve as a viable defense strategy against CDA. In the kernel space, defenses such as Supervisor Mode Execution Protection (SMEP) and Supervisor Mode Access Prevention (SMAP)~\cite{smap} already prevent kernel-mode code from directly accessing user-mode memory, effectively isolating different privilege levels. A similar mechanism could be adopted in the hypervisor setting: the host would be restricted from accessing guest memory by default, with hardware enforcing a strong access barrier between host and guest.

Such isolation would prevent unintended or malicious host interactions with guest memory, significantly reducing the risk of CDA exploits that rely on redirecting host pointers into guest-controlled regions. However, certain operations—such as device emulation or management tasks—legitimately require the host to access guest memory. To support these needs securely, access could be selectively permitted through dedicated APIs or hardware instructions that temporarily relax the isolation under tightly controlled conditions. Inspired by the \texttt{STAC} and \texttt{CLAC} instructions used by SMAP, hypervisors could adopt similar mechanisms to grant temporary, auditable access to guest memory only when explicitly invoked.

\textbf{Gadget Reduction}. An alternative line of defense focuses on reducing the presence of cross-domain gadgets. To achieve this, hypervisors should avoid storing raw guest-HVA pointers in host memory. Instead, guest memory references can be represented using opaque tokens such as handles or offsets, have no direct dereference semantics. When memory access is required, these tokens are resolved through a secure lookup mechanism to obtain the corresponding HVA. Another complementary approach is to cryptographically encode guest-HVA pointers, similar to pointer authentication (PAC), before storing them, and to decode them only upon verified use. This ensures that no valid guest address remains in cleartext within host memory, thereby preventing potential misuse. These designs eliminate the presence of dereferenceable guest addresses in host memory, thereby blocking a key prerequisite for CDA-based exploitation.

\subsection{Potential Existence of Additional Gadgets}

Following a best-effort approach, we collected translation functions from official documentation and validated them through static analysis. However, undocumented routines may still exist and inadvertently retain guest-HVA pointers, forming implicit cross-domain gadgets beyond our current coverage. While our analysis captures the major patterns observed in practice, we acknowledge that additional cases could emerge and warrant further investigation.

\subsection{Limitations on Confidential VMs}

While our CDA framework demonstrates effective cross boundary exploitation in traditional virtualized environments, its applicability to Confidential Virtual Machines (CVMs), such as those protected by AMD SEV-SNP or Intel TDX, is limited. These platforms enforce strict memory encryption and isolation: guest memory and CPU state are encrypted and integrity protected, so that the hypervisor or host cannot directly read or modify guest data. As a result, key primitives required by CDA will be restricted.

However, shared pages remain accessible in plaintext to the host, such as GHCB pages, TDX shared buffers, and virtio shared rings~\cite{cvm_shared}. If the hypervisor does not properly validate and constrain its use of these shared regions, CDA-style attacks may still reappear, since the attacker fully controls these buffers. Overall, although CVMs significantly reduce the gadget surfaces and translation visibility needed by CDA, the fundamental idea of cross-domain pointer manipulation remains relevant and merits further exploration in isolation-enhanced environments.

\subsection{Applicability to KVM Integration}

Within the QEMU virtualization setting considered in this work, our scope is confined to the userspace portion, where guest memory is mapped into the process address space and exposed as host-side HVAs. At the same time, we note that the KVM subsystem (Kernel-based Virtual Machine~\cite{kvm}) also maintains HVA pointers as part of its normal operation. Although this work does not examine whether similar CDA-like behaviors could arise at the kernel boundary, we do not rule out this possibility. A systematic investigation of cross-domain pointer interactions within pure KVM or other kernel-level virtualization paths remains promising future work and may reveal additional opportunities for understanding CDA beyond the userspace components of the hypervisor.

\subsection{Applicability to Other Scenarios}

The core concept of CDA revolves around exploiting the weak isolation between two distinct memory spaces. By taking advantage of this weak separation, an attacker can manipulate the memory layout of the target space by using the controllable environment of their own space, thereby facilitating successful exploitation. This approach is particularly dangerous because it leverages the interaction between two spaces that lack rigorous boundary validation, making a wide range of systems vulnerable to this type of attack. Essentially, any environment where two isolated memory spaces interact—such as Software Guard Extensions (SGX), Trusted Execution Environments (TEE), microservices architectures, and inter-process communication systems—can be at risk if they do not enforce strict boundary checks. The concept becomes especially effective in minimalistic or constrained architectures, where available primitives are limited, but one space remains fully controllable by the attacker. In such scenarios, where a weak boundary exists between the interacting domains, an attacker can manipulate one domain to influence the other, bypassing conventional security defenses. The potential impact of CDA-style attacks is broad, extending to any system where domain separation is not rigorously maintained. This underscores the critical need for robust boundary validation and strict isolation policies to protect against such sophisticated attacks, as any weakness in these areas could provide an entry point for exploitation.

%% file: 0x07RelatedWork.tex
% \vspace{-2mm}
\section{Related Work}
\textbf{Hypervisor Vulnerability Discovery.} In the field of virtualization, vulnerability discovery has always been a focal point for security researchers. Early efforts began with dump fuzzing~\cite{xenfuzz,viridian,gorobets2015attacking}, followed by advancements in addressing data input issues~\cite{pan2021v,bulekov2022morphuzz,schumilo2020hyper,henderson2017vdf}, and later incorporating generation of high-quality test case sequences~\cite{myung2022mundofuzz,liu2023videzzo,liu2023vd,zhang2025insvdf} and hardware-assisted methods~\cite{schumilo2021nyx,bulekov2024hyperpill}. For example, Hyper-Cube~\cite{schumilo2020hyper} utilizes a custom operating system to achieve high-throughput, multi-dimensional fuzzing. Nyx~\cite{schumilo2021nyx} employs full-system snapshots and a hardware-assisted coverage framework to fuzz hypervisors. V-Shuttle~\cite{pan2021v} uses DMA redirection to flatten nested structures, while MORPHUZZ~\cite{bulekov2022morphuzz} reshapes the input space of virtual devices by collecting feedback from the core hypervisor's interface APIs. ViDeZZo~\cite{liu2023videzzo} leverages static analysis to extract dependencies within and between messages. Over time, virtualization vulnerability discovery techniques have made significant progress. However, vendors are increasingly focused on whether these vulnerabilities can be exploited in cloud environments~\cite{kvmctf}.

\textbf{Kernel Exploitation.} The development of kernel exploitation techniques has progressed through several stages, gradually forming a rich methodological framework. Early research primarily focused on bypassing various kernel protections to ensure successful exploitation (e.g., ~\cite{hund2009return,kemerlis2014ret2dir}). As the field matured, the research focus shifted toward generating exploits for specific vulnerabilities, exemplified by tools like Collision, Fuze, Koobe, and Exprace~\cite{xu2015collision,wu2018fuze,chen2020koobe,lee2021exprace,wu2019kepler,lu2017unleashing,cho2020exploiting}, driving significant advancements in the precision and stability of exploit generation. Building on this foundation, further research has focused on enhancing the stability of heap exploitation~\cite{chen2019slake,zeng2022playing} and actively exploring new exploitation primitives~\cite{yun2020automatic,chen2020systematic,koschel2023uncontained,wang2023alphaexp,avllazagaj2024scavy}. New exploitation methods continue to emerge, such as the DirtyCred and PSpray techniques~\cite{lin2022dirtycred,lee2023pspray}, injecting fresh perspectives into kernel exploitation. Meanwhile, the advent of the Automated Exploit Generation (AEG) technology has further increased automation~\cite{heelan2018automatic,wang2021maze}, making exploitation more efficient and accessible. These advancements indicate that kernel exploitation has now established a solid technical foundation and a mature toolkit. By comparison, exploitation in virtualization systems remains far less developed. Although cross-domain ideas such as ret2usr~\cite{kemerlis2014ret2dir} show how kernels can redirect execution into user memory, these techniques do not translate to hypervisors, which lack direct access to guest memory and must instead rely on GPA-to-HVA translation. CDA is novel in that it systematically reveals when hypervisors unintentionally reintroduce guest-HVA pointers into host memory, formalizes these behaviors into four semantic variants, and turns guest memory into a practical, reusable exploitation substrate.

\textbf{Hypervisor Exploitation.} 
Current VM-to-hypervisor exploitation research has produced numerous practical VM-escape attacks across both open-source and proprietary hypervisors—including QEMU~\cite{elhage2011virtunoid,shao20203d,pan2021scavenger,venom,limatryoshka}, VirtualBox~\cite{VBescape,VBpwn2own}, VMware~\cite{speedpwn,greatescapes,VMware-uhci,a-ctf-style,URBExcalibur}, and ESXi~\cite{zhao2019breaking}. These attacks span virtual devices, network stacks, and shared services and primarily rely on identifying exploitable host-side structures such as function pointers or corrupted objects. As a result, prior techniques remain highly ad-hoc, heavily dependent on expert-crafted primitives, and tied to specific host-resident artifacts. Moreover, existing work is almost entirely \textbf{host-centric}, treating guest memory only as attacker input rather than as a first-class exploitation substrate. Although prior studies (e.g.,~\cite{pan2021scavenger}) showed isolated cases where hypervisors may dereference guest-controlled addresses, they did not characterize the underlying conditions, generality, or semantic variants of such cross-domain behaviors. \textbf{CDA departs fundamentally from these earlier VM-to-hypervisor attacks} by systematizing this phenomenon: we identify the root causes of cross-domain pointer flows, formalize four canonical variants based on pointer-use semantics, and demonstrate that these opportunities arise broadly across hypervisor codebases. This work therefore elevates CDA from scattered empirical observations to a \textbf{general, structured exploitation paradigm}—a perspective not captured in prior literature.

%% file: 0x08Conclusion.tex
% \vspace{-0.5em}
\section{Conclusion}

In virtualization environments, exploiting pointer corruption vulnerabilities is particularly challenging due to the scarcity of exploitable data structures in the host. Traditional techniques often lack a stable foothold for redirecting corrupted pointers, limiting their effectiveness. In this paper, we provide the first systematic characterization of CDA, which leverages weak guest-host isolation to redirect corrupted host pointers to attacker-controlled payloads in guest memory. Our findings reveal that cross-domain attacks pose a significant and emerging threat to the security of existing virtualization infrastructures. This transforms guest memory into a reusable and controllable primitive for exploitation. Additionally, we develop an automated system to identify cross-domain gadgets and construct exploitation chains, demonstrating the approach’s practicality across real-world hypervisors. We hope this work raises awareness in the virtualization and cloud security communities. We hope this work raises awareness in the virtualization and cloud security communities about the risks of implicit trust in guest memory and motivates stronger isolation mechanisms.

%% file: 0x09Appendix.tex
\section{Address Translation Functions}
\label{sec:translation}

Table~\ref{tab:translation_functions} lists the hypervisor functions responsible for guest physical address (GPA) to host virtual address (HVA) translation across two widely used virtualization platforms. In QEMU, this functionality is implemented by functions such as \texttt{address\_space\_map()} and other variants, which manage memory mapping and access for guest memory regions. In VirtualBox, similar roles are performed by functions such as \texttt{PGMPhysWrite()} and \nolinkurl{PGMR3PhysBulkGCPhys2CCPtrExternal()}, which provide fine-grained control over guest-to-host address resolution and data access.

\begin{table}[h]
\centering
\resizebox{\linewidth}{!}{%
\begin{tabular}{@{}ll@{}}
\toprule
Category                    & Translation Functions                        \\ \midrule
\multirow{5}{*}{QEMU}       & address\_space\_map();                       \\
                            & address\_space\_unmap();                     \\
                            & address\_space\_read\_full();                \\
                            & address\_space\_read();                      \\
                            & address\_space\_write();                     \\ \midrule
\multirow{4}{*}{Virtualbox} & PGMPhysWrite(); PGMPhysRead();               \\
                            & PGMR3PhysBulkGCPhys2CCPtrExternal();         \\
                            & PGMR3PhysBulkGCPhys2CCPtrReadOnlyExternal(); \\
                            & PGMR3PhysGCPhys2CCPtrExternal();             \\ \bottomrule
\end{tabular}%
}
\caption{Hypervisor address translation functions.}
\label{tab:translation_functions}
\end{table}

\begin{figure*}
    \centering
    \begin{subfigure}[b]{0.48\textwidth}
    \includegraphics[width=1\linewidth]{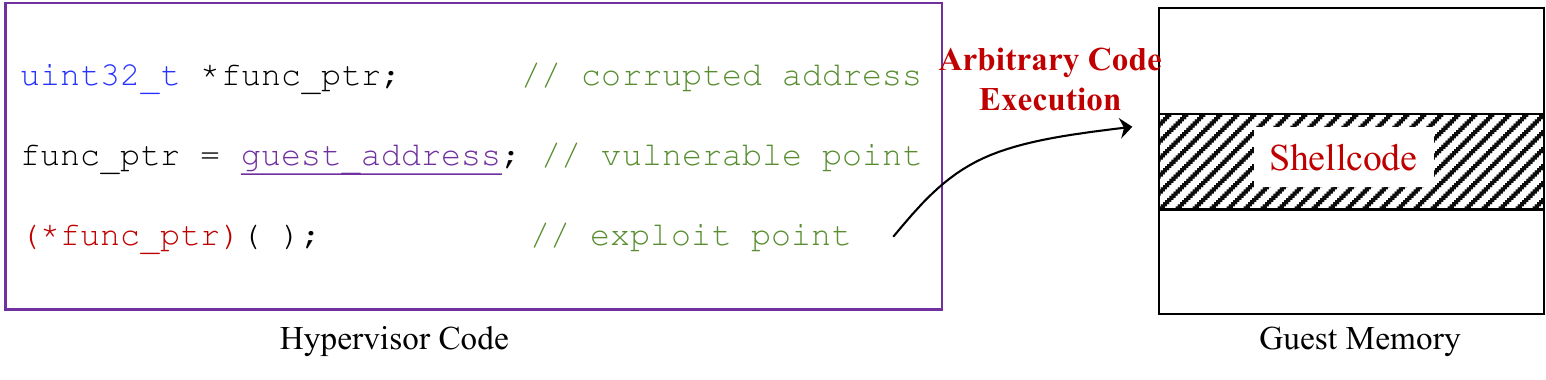}
    \caption{Example illustrating how CDA causes arbitrary code execution.}
    \label{fig:variant1}
    \end{subfigure}
    \hspace{4mm}
    \begin{subfigure}[b]{0.48\textwidth}
    \includegraphics[width=1\linewidth]{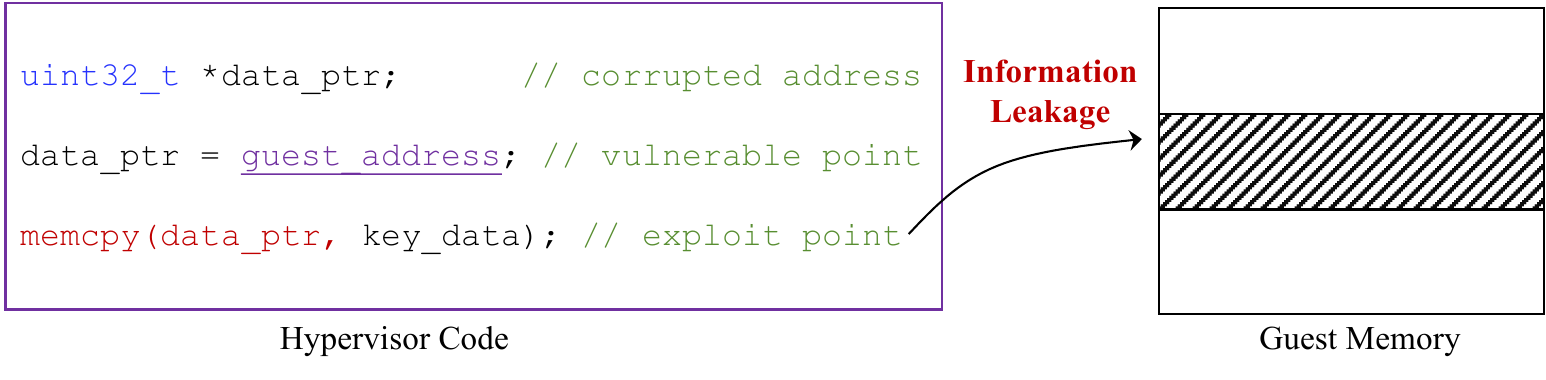}
    \caption{Example illustrating how CDA can leak information.}
    \label{fig:variant2}
    \end{subfigure}

    \begin{subfigure}[b]{0.48\textwidth}
    \includegraphics[width=1\linewidth]{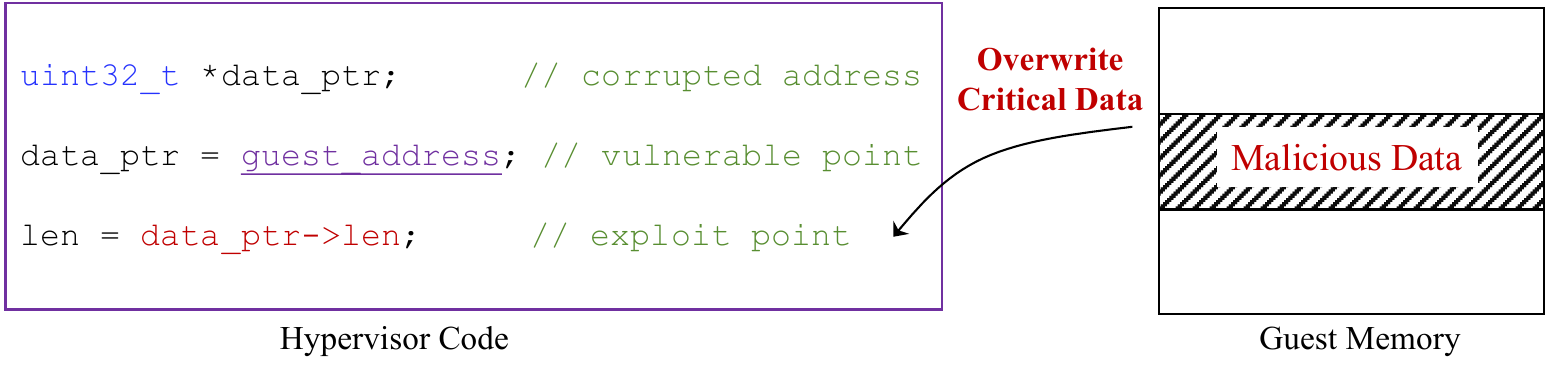}
    \caption{Example illustrating how CDA can overwrite critical data.}
    \label{fig:variant3}
    \end{subfigure}
    \hspace{4mm}
    \begin{subfigure}[b]{0.48\textwidth}
    \includegraphics[width=1\linewidth]{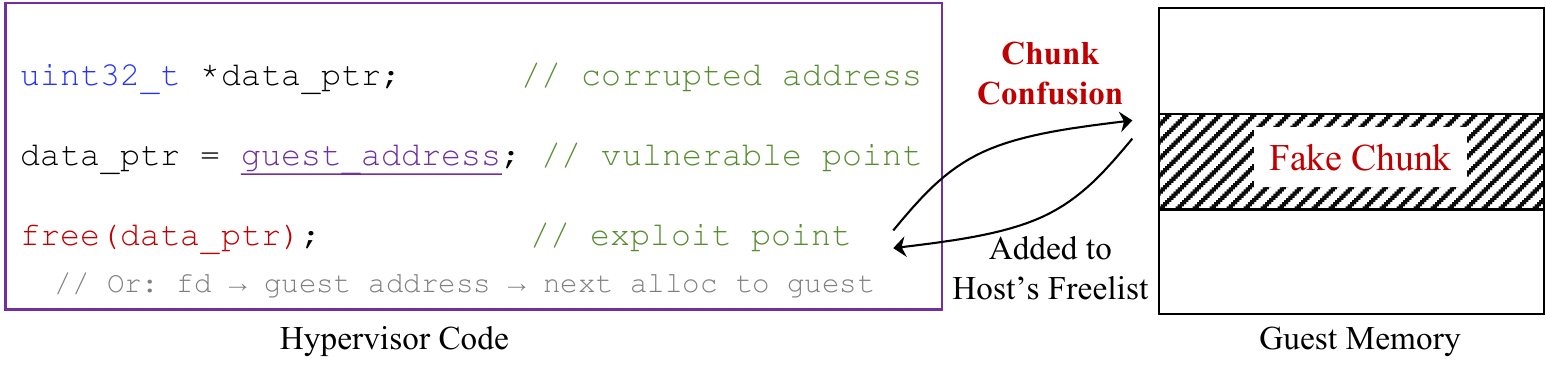}
    \caption{Example illustrating how CDA can craft chunk confusion.}
    \label{fig:variant4}
    \end{subfigure}
    
    \caption{CDA variants to exploit pointer corruption vulnerabilities.}
    \label{fig:vatiants}
\vspace{-1em}
\end{figure*}

\section{CDA Variants}
\label{sec:variants}

We summarize four variants of CDA, categorized by the usage context of the corrupted pointer—namely when the vulnerable code performs a \textbf{free}, \textbf{call}, \textbf{read}, or \textbf{write} operation on it—which collectively cover all possible exploitation effects of CDA, as detailed below:

\subsubsection{Arbitrary Code Execution (CDA\textsuperscript{\textbf{A}})}
We begin by introducing the first CDA variant: arbitrary code execution, as illustrated in Figure~\ref{fig:variant1}. This variant targets cases where a pointer corruption vulnerability results in a corrupted function pointer, which is subsequently dereferenced by the hypervisor. Instead of attempting to locate a valid function address within the host—an approach made difficult by ASLR, limited code reuse gadgets—CDA takes a more flexible path by redirecting the corrupted pointer to guest memory. Guest memory, while isolated from the guest’s perspective, is still directly accessible by the host. Thus, once the corrupted function pointer is invoked, the control flow is transparently transferred to an attacker-controlled region in guest space, where a staged shellcode resides. This effectively bypasses the need for reliable host-side code reuse primitives or shellcode injection mechanisms.

\subsubsection{Information Leakage (CDA\textsuperscript{\textbf{I}})}
Next, we introduce the second CDA variant: sensitive information leakage, as illustrated in Figure~\ref{fig:variant2}. This scenario arises when a pointer corruption vulnerability leaves behind a corrupted data pointer that is later used in a write operation. Instead of pointing to a legitimate buffer in host memory, CDA redirects this pointer to an attacker-controlled address within the guest memory space. As a result, when the hypervisor writes data to the corrupted pointer, it unintentionally writes directly into guest memory—effectively leaking sensitive information from the host to the guest. This variant enables a side-effect-free and stealthy data exfiltration channel, without requiring any interaction through standard I/O mechanisms such as DMA buffers or MMIO transactions. Traditional interfaces often impose strict constraints, such as requiring pre-allocated transfer objects, matching data formats, or bounded transfer sizes. CDA bypasses these limitations and eliminates the need for additional communication logic.

\subsubsection{Critical Data Overwriting (CDA\textsuperscript{\textbf{O}})}
We now present the third CDA variant: critical data overwriting, as illustrated in Figure~\ref{fig:variant3}. In this scenario, a pointer corruption vulnerability leaves behind a corrupted data pointer that is later used in a read operation. CDA redirects this pointer to guest memory, causing the hypervisor to read attacker-controlled data directly from the guest space. As a result, the guest-controlled input is implicitly trusted and used to update sensitive fields within the hypervisor. This variant effectively allows the attacker to inject arbitrary values into critical data structures without the need for traditional write primitives. Since the guest space is entirely under the attacker’s control, the attacker can craft malicious data with fine-grained control over content, layout, and size, making the overwrite both precise and stealthy. By abusing the host’s implicit trust in the source of a read operation, this variant converts a seemingly benign data corruption into a powerful logic corruption primitive, enabling further exploitation such as privilege manipulation, control flag flipping, or fake object injection.

\subsubsection{Chunk Confusion (CDA\textsuperscript{\textbf{C}})}
Finally, we describe the last CDA variant: chunk confusion, as illustrated in Figure~\ref{fig:variant4}. This attack occurs when the corrupted address produced by a pointer corruption vulnerability is involved in memory allocation or deallocation, leading to inconsistencies in the hypervisor's heap management logic. We formalize two representative scenarios:

(1) If the corrupted address is used in a \texttt{free()} operation, CDA redirects it to guest memory, causing the hypervisor to erroneously free an attacker-crafted fake chunk residing in the guest space. This fake chunk is then inserted into the host’s freelist and treated as a valid memory region by the host-side allocator. As a result, future memory allocation requests from the hypervisor may return pointers to guest memory, allowing the attacker to gain full read/write access to sensitive host data structures allocated on top of it. This effectively enables a guest-assisted heap spraying primitive, leveraging the hypervisor’s own memory manager.

(2) Similarly, when the corrupted value affects heap metadata—such as a manipulated \texttt{fd} pointer in a free chunk—CDA can redirect the allocator’s next chunk candidate into guest space. During the next allocation, the hypervisor mistakenly allocates memory from the attacker-controlled guest region. This allows the guest to pre-position payloads or metadata that will later be interpreted as legitimate host structures, enabling logic corruption or type confusion attacks without violating memory access protections.

These two patterns demonstrate how CDA can subvert the hypervisor’s memory allocator by injecting crafted objects from the guest domain, enabling precise control over the heap layout and allocation behavior in host space. Unlike traditional heap exploits that require complex heap feng shui in host memory, CDA provides a lightweight and deterministic alternative.

\begin{figure}[t]
\centering
\begin{lstlisting}
int virtio_gpu_create_mapping_iov(VirtIOGPU *g, uint32_t nr_entries, struct iovec **iov)
{
    *iov = g_malloc0(sizeof(struct iovec) * nr_entries);
    for (i = 0; i < nr_entries; i++) {
        uint64_t addr = ents[i].addr;
        uint32_t len = ents[i].length;
        (*iov)[i].iov_len = len;
        (*iov)[i].iov\_base = (*@\hl{dma\_memory\_map(addr, \&len);}@*)
    }
    return 0;
}
\end{lstlisting}
\vspace{-0.5em}
\caption{
A elastic gadget in \texttt{virtio\_gpu\_create\_mapping\_iov} that calls \texttt{dma\_memory\_map} with a guest-provided address \texttt{addr} and length \texttt{len}. 
The return value is stored in \texttt{iov\_base}, forming a host-accessible pointer derived from guest input.
}
\label{fig:qemu_gpu_dma_gadget}
\end{figure}

\section{Example of Elastic Cross-Domain Gadget}
\label{sec:elastic}

To illustrate the versatility of elastic cross-domain gadgets, we present a representative example from the virtio-GPU device. As shown in Figure~\ref{fig:qemu_gpu_dma_gadget}, the device allocates a data structure whose size is directly controlled by the guest (line 3). This structure contains an array of \texttt{iovec} entries, each embedding a length field (line 7) and a guest physical address (line 8). The resulting layout forms a one-to-one mapping table of guest addresses, with entries spaced at regular 0x10-byte intervals.

Because both the total allocation size and the number of embedded guest addresses are determined by the guest, this layout acts as a highly flexible \textbf{elastic cross-domain gadget}, capable of adjusting its size, alignment, and pointer positions to match diverse heap-based exploitation needs. Importantly, similar elastic patterns emerge across multiple virtio backends, indicating that this capability is inherent to their allocation and structure design rather than a device-specific anomaly.

\section{Detailed Cross-Domain Gadgets}

To support our analysis of CDA, we provide a comprehensive breakdown of gadget instances across various device subsystems. Table~\ref{tab:cross_domain_gadgets} summarizes 776 gadget instances identified in our study, categorized by gadget family, trigger function pair, and their guest-controlled GPA source field.

\begin{table*}[t]
\centering
\caption{Summary of cross-domain gadgets in the QEMU: Each gadget is triggered by a guest-controllable field and mapped to host address space via \texttt{MMIO} or \texttt{Timer/BH} mechanisms.}
\label{tab:cross_domain_gadgets}
\resizebox{\textwidth}{!}{
\begin{tabular}{@{}lllcccl@{}}
\toprule
\textbf{Gadget Family} & \textbf{Upper Function} & \textbf{Translation Function} & \textbf{HVA Variable} & \textbf{GPA Source Field} & \textbf{Trigger Type} & \textbf{Count} \\
\midrule
\multirow{4}{*}{DMA gadget}
 & dma\_memory\_map        & address\_space\_map     & ad$\rightarrow$lst & AHCIPortRegs$\rightarrow$fis\_addr         & MMIO (4), BH (0)    & 4 \\
 & pci\_dma\_map           & address\_space\_map     & ring$\rightarrow$page & txd.addr & MMIO (10), BH (0)   & 10 \\
 & dma\_memory\_read       & address\_space\_read\_full    & ram\_ptr  & s$\rightarrow$tx\_descriptor                & MMIO (81), BH (21)  & 102 \\
 & dma\_memory\_write      & address\_space\_write   & ram\_ptr & s$\rightarrow$tx\_descriptor                   & MMIO (91), BH (16)  & 107 \\
\midrule
\multirow{6}{*}{USB gadget}
 & usb\_packet\_map        & address\_space\_map     & packet$\rightarrow$iov & sgl$\rightarrow$sg[num\_sg].base & MMIO (3), BH (11)   & 14 \\
 & get\_dwords             & address\_space\_read\_full    & ram\_ptr & q$\rightarrow$qhaddr & MMIO (1), BH (35)   & 36 \\
 & put\_dwords             & address\_space\_write   & ram\_ptr & q$\rightarrow$qhaddr & MMIO (2), BH (44)   & 46 \\
 & xhci\_dma\_read\_u32s   & address\_space\_read\_full    & ram\_ptr & sctx$\rightarrow$pctx          & MMIO (22), BH (8)   & 30 \\
 & xhci\_dma\_write\_u32s  & address\_space\_write   & ram\_ptr & sctx$\rightarrow$pctx           & MMIO (17), BH (4)   & 21 \\
 & xhci\_write\_event      & address\_space\_write   & ram\_ptr & intr$\rightarrow$er\_start & MMIO (17), BH (5)   & 22 \\
\midrule
\multirow{2}{*}{Virtio gadget}
 & virtqueue\_map\_desc    & address\_space\_map     & iov[num\_sg].iov\_base & desc[num\_sg].addr         & MMIO (40), BH (16)  & 56 \\
 & virtio\_gpu\_create\_mapping\_iov    & address\_space\_map     & iov[num\_sg].iov\_base & desc[num\_sg].addr         & MMIO (0), BH (2)  & 2 \\
\midrule
Display gadget
 & cpu\_physical\_memory\_map & address\_space\_map  & data & s$\rightarrow$dispc.l[0].addr[0] & MMIO (1), BH (0)    & 1 \\
\midrule
Block device gadget
 & dma\_blk\_cb            & address\_space\_map     &  dbs$\rightarrow$iov & req$\rightarrow$sg.qsg  & MMIO (4), BH (6)    & 10 \\
\midrule
\multirow{2}{*}{SCSI gadget}
 & lsi\_mem\_read          & address\_space\_read    & ram\_ptr & s$\rightarrow$dsp            & MMIO (4), BH (0)    & 4 \\
 & lsi\_mem\_write         & address\_space\_write   & ram\_ptr & s$\rightarrow$dsp           & MMIO (4), BH (0)    & 4 \\
\midrule
\multirow{2}{*}{PCI gadget}
 & pci\_dma\_read          & address\_space\_read\_full    & ram\_ptr & r$\rightarrow$bdbar                 & MMIO (59), BH (120) & 179 \\
 & pci\_dma\_write         & address\_space\_write   & ram\_ptr & desc.buffer\_addr                & MMIO (42), BH (22)  & 64 \\
\midrule
\multirow{3}{*}{SDHCI gadget}
 & sdhci\_do\_adma         & address\_space\_read\_full    & ram\_ptr & dscr.addr   & MMIO (12), BH (2)   & 14 \\
 & sdhci\_sdma\_transfer\_multi\_blocks       & address\_space\_read\_full    & ram\_ptr & s$\rightarrow$sdmasysad   & MMIO (9), BH (1)    & 10 \\
 & sdhci\_sdma\_transfer\_multi\_blocks      & address\_space\_write   & ram\_ptr & desc.buffer\_addr   & MMIO (9), BH (1)    & 10 \\
\hline
\multicolumn{6}{l}{\textbf{Total}} & \textbf{772} \\
\bottomrule
\end{tabular}
}
\end{table*}